\documentclass[prd,aps,nofootinbib,superscriptaddress,preprintnumbers,10pt]{revtex4-2}
\usepackage{mathrsfs}
\usepackage{dsfont}
\usepackage{csquotes}

\usepackage{amsmath, amssymb, amsfonts}
\usepackage{mathtools}
\usepackage{amsthm}
\usepackage{siunitx}
\usepackage{slashed}      
\usepackage{bm}           
\usepackage{empheq}

\usepackage{graphicx}
\usepackage[percent]{overpic}
\usepackage{adjustbox}
\usepackage[section]{placeins}     
\usepackage{epsfig}

\usepackage{booktabs}
\usepackage{array}       
\usepackage{tablefootnote}
\usepackage{hhline}

\usepackage{enumitem}

\usepackage{listings}

\usepackage{standalone}
\usepackage{import}
\usepackage{comment}
\usepackage{appendix}     
\usepackage{ifsym}        
\usepackage[normalem]{ulem} 

\usepackage[table,xcdraw]{xcolor} 
\usepackage[most]{tcolorbox}      
\usepackage{soul}

\usepackage{tikz}
\usetikzlibrary{matrix,fit,decorations.markings}
\usepackage[compat=1.1.0]{tikz-feynman}
\usepackage{silence}
\usepackage{simplewick}
\usepackage{simpler-wick}

\usepackage[unicode,colorlinks=true,linkcolor=black,citecolor=black,urlcolor=blue]{hyperref}
\hypersetup{pdfstartview=FitH,bookmarksopen=true,pdfborder={0 0 0}}

\usepackage[capitalise]{cleveref}

\renewcommand{\arraystretch}{1.5}
\newcommand{\Ttabular}{\rule{0pt}{4ex}}
\newcommand{\Btabular}{\rule[-2.5ex]{0pt}{2.5ex}}

\definecolor{DarkBlue}{HTML}{003366}
\definecolor{BurntOrange}{HTML}{CC5500}
\definecolor{MyGreen}{HTML}{005035}
\definecolor{MyLightGreen}{HTML}{C3D7A4}
\definecolor{MyGold}{HTML}{A49665}
\definecolor{MyPink}{HTML}{DE3A6E}
\definecolor{White}{HTML}{FFFFFF}
\definecolor{BrickRed}{HTML}{993333}
\definecolor{Mauve}{HTML}{A76B88}
\definecolor{pine}{rgb}{0.0, 0.5, 0.0}

\newcommand\n[1]{{:}\mkern1mu#1\mkern1.6mu{:}}

\newcommand{\pdagg}[0]{{\phantom{\dagger}}}
\DeclarePairedDelimiter\bra{\langle}{\rvert}
\DeclarePairedDelimiter\ket{\lvert}{\rangle}

\DeclarePairedDelimiterX\braket[2]{\langle}{\rangle}{#1\,\delimsize\vert\,\mathopen{}#2}
\renewcommand{\[}{\begin{equation}}
\renewcommand{\]}{\end{equation}}
\newcommand{\ta}{\left(}
\newcommand{\tc}{\right)}
\newcommand{\qa}{\left[}
\newcommand{\qc}{\right]}

\newcommand{\diagram}[2][0.8]{\adjustbox{scale=#1, valign=c}{\includestandalone{Diagrams/#2}}}
\newcommand\scalemath[2]{\scalebox{#1}{\mbox{\ensuremath{\displaystyle #2}}}} 

\definecolor{SRmauve}{rgb}{0.58,0,0.82}

\tikzset{W->-/.style={decoration={
  markings,
  mark=at position 0.5*\pgfdecoratedpathlength+2pt with
  {\draw[-latex] (-2pt,0pt) -- (1pt,0pt);}},postaction={decorate}},
  W-<-/.style={decoration={
  markings,
  mark=at position 0.5*\pgfdecoratedpathlength with
  {\draw[latex-] (-2pt,0pt) -- (1pt,0pt);}},postaction={decorate}}
}

\pgfkeys{
  /simplerwick/.cd,
  arrows/.store in=\LstWickArrows,
  arrows/.initial={-,-,-,-,-,-,-,-,-}, 
  arrows={-,-,-,-,-,-,-,-,-},
  positions/.store in=\LstWickPositions,
  positions={+1,+1,+1,+1,+1,+1,+1,+1,+1},
  positions/.initial={+1,+1,+1,+1,+1,+1,+1,+1,+1},
}

\makeatletter
\newcounter{Wick@up}
\newcounter{Wick@down}
\def\swick@end#1#2{
  \swick@setfalse@#1
  \tikzexternaldisable
  \begin{tikzpicture}[remember picture, baseline=(swick-close#1.base)]
    \node[use as bounding box, inner sep=0pt, outer sep=0pt] (swick-close#1) {$\displaystyle #2$};
  \end{tikzpicture}
  \tikz[remember picture, overlay]
{
\xdef\myW@style{\empty}
\foreach \W@X[count=\W@C] in \LstWickArrows
{\ifnum\W@C=#1
\xdef\myW@style{\W@X}
\fi}
\ifx\myW@style\empty
\PackageWarning{simpler-wick}{%
The list arrows has not enough entries!%
}{}
\xdef\myW@style{-}
\fi
\xdef\myW@pos{-77}
\foreach \W@X[count=\W@C] in \LstWickPositions
{\ifnum\W@C=#1
\xdef\myW@pos{\W@X}
\fi}
\ifnum\myW@pos=-77
\PackageWarning{simpler-wick}{%
The list positions has not enough entries!%
}{}
\xdef\myW@pos{+1}
\fi
\ifnum\myW@pos=-1
    \draw[\myW@style] ($(swick-open#1.south) + (0, -3pt)$) 
          -- ($(swick-open#1.base) + (0, -\swick@offset) + \theWick@down*(0, -\swick@sep)$) 
          -- ($(swick-close#1.base) + (0, -\swick@offset) + \theWick@down*(0, -\swick@sep)$) 
          -- ($(swick-close#1.south) + (0, -3pt)$);
\stepcounter{Wick@down}
\else
\stepcounter{Wick@up}
    \draw[\myW@style] ($(swick-open#1.north) + (0, 3pt)$) 
          -- ($(swick-open#1.base) + (0, \swick@offset) + \theWick@up*(0, \swick@sep)$) 
          -- ($(swick-close#1.base) + (0, \swick@offset) + \theWick@up*(0, \swick@sep)$) 
          -- ($(swick-close#1.north) + (0, 3pt)$);
\fi}
  \tikzexternalenable}
\def\wick@[#1]#2{\setcounter{Wick@up}{0}
\setcounter{Wick@down}{-1}
  \ifmmode
    \begingroup
    \pgfkeys{
        simplerwick,
        #1}
    \swick@cond@reset
    \swick@count=0
    \def\swick@max{0}
    \def\c{\swick@smart}
    #2
    \dimen0=\swick@sep
    \multiply\dimen0 by \swick@max
    \advance\dimen0 by \swick@offset
    \vbox to \dimen0{}
    \swick@cond@any{
      \PackageWarning{simpler-wick}{%
        I have reached the end of \protect\wick\space with some unclosed
        contractions%
      }{}
    }{}
    \endgroup
  \else
    \PackageWarning{simpler-wick}{%
      \protect\wich\space has been called outside a math environment, this will
      be ignore%
    }
  \fi
}
\makeatother

\allowdisplaybreaks

\makeatletter
    \def\CT@@do@color{%
      \global\let\CT@do@color\relax
            \@tempdima\wd\z@
            \advance\@tempdima\@tempdimb
            \advance\@tempdima\@tempdimc
    \advance\@tempdimb\tabcolsep
    \advance\@tempdimc\tabcolsep
    \advance\@tempdima2\tabcolsep
            \kern-\@tempdimb
            \leaders\vrule
                    \hskip\@tempdima\@plus  1fill
            \kern-\@tempdimc
            \hskip-\wd\z@ \@plus -1fill }
    \makeatother

\begin{document}

\title{Higher-order structure of Hamiltonian truncation effective theory}

\author{Andrea Maestri}
\thanks{Electronic address: \href{mailto:andrea.maestri01@universitadipavia.it}{andrea.maestri01@universitadipavia.it}}
\affiliation{Dipartimento di Fisica ``A. Volta'', Universit\`a degli Studi di Pavia, I-27100 Pavia, Italy}
\affiliation{Istituto Nazionale di Fisica Nucleare, Sezione di Pavia, I-27100 Pavia, Italy}

\author{Simone Rodini}
\thanks{Electronic address: \href{mailto:simone.rodini@unipv.it}{simone.rodini@unipv.it}}
\affiliation{Dipartimento di Fisica ``A. Volta'', Universit\`a degli Studi di Pavia, I-27100 Pavia, Italy}
\affiliation{Istituto Nazionale di Fisica Nucleare, Sezione di Pavia, I-27100 Pavia, Italy}

  \author{Barbara Pasquini}
\thanks{Electronic address:  \href{mailto:barbara.pasquini@unipv.it}{barbara.pasquini@unipv.it} }
\affiliation{Dipartimento di Fisica ``A. Volta'', Universit\`a degli Studi di Pavia, I-27100 Pavia, Italy}
\affiliation{Istituto Nazionale di Fisica Nucleare, Sezione di Pavia, I-27100 Pavia, Italy}

\date{\today}
\begin{abstract}
We study the Hamiltonian truncation for the two-dimensional $\lambda\phi^4$ theory within the framework of Hamiltonian truncation effective theory, where truncation artifacts are mitigated through a systematic inclusion of corrective terms organized in inverse powers of the ultraviolet energy cut-off $E_{\rm max}$. Building on the leading-order matching program, we develop two complementary extensions. First, we derive compact all-order expressions for the local matching corrections to the mass and quartic coupling by resumming infinite classes of diagrams sharing fixed topologies within the local approximation. Second, we extend the non-local sector by computing the next-to-next-to-local corrections contributing at $\mathcal{O}(E_{\rm max}^{-4})$, following a continuum-first matching procedure, in which the effective corrections are computed in infinite volume and the spatial direction is subsequently re-compactified to obtain a discrete basis of free-Hamiltonian eigenstates on which the truncated operator construction is implemented. Our results show that an increasingly rich operator basis is necessary to describe the theory beyond leading order.
\end{abstract}
\maketitle

\section{Introduction}
\label{sec_Introduction}
Quantum field theory (QFT) provides the conceptual framework underlying the Standard Model, whose perturbative predictions have been tested with high precision across a wide range of energies. Many phenomena of interest in fundamental and many-body systems are, however, intrinsically non-perturbative. Some examples include confinement and the generation of mass gaps in non-abelian gauge theories, real-time dynamics in strongly correlated or out-of-equilibrium settings, and strongly coupled renormalization group flows~\cite{Wilson1972,Wilson:1974sk,Berges:2004yj,Wetterich:1992yh}.
 In such regimes, the canonical perturbative expansion, for instance in terms of Feynman diagrams, is not reliable, and non-perturbative methods are required.

A standard non-perturbative route is the Euclidean lattice path-integral formulation, whose Monte Carlo evaluation has enabled first-principles results for a broad class of observables in strongly interacting theories~\cite{Creutz:1980zw}. Lattice Monte Carlo, however, is not universally applicable. In particular, finite-density systems and many real-time observables are affected by sign problems, which significantly restrict the practical use of importance-sampling techniques~\cite{deForcrand:2009zkb,Gattringer:2016kco}. These considerations have prompted the development of complementary approaches that are formulated directly in a Hamiltonian and/or real-time framework, including the Hamiltonian lattice formulation~\cite{Kogut:1974ag}, tensor-network methods~\cite{Banuls:2019rao,Silvi:2014pta}, and quantum-simulation inspired strategies~\cite{Kaplan:2018vnj,Brower:2020huh}.

Within this context, Hamiltonian truncation (HT) methods have attracted renewed attention. In these approaches, the full QFT Hamiltonian is approximated by restricting it to a finite-dimensional subspace of the Hilbert space and subsequently diagonalizing the resulting matrix.
This approach originates in the truncated conformal space approach, introduced by Yurov and Zamolodchikov~\cite{Yurov1990,Yurov1991}, and it has since been developed along several directions. These include applications to a broad class of two-dimensional quantum field theories, such as the Ising model, the sine-Gordon and massive Schwinger models, as well as gauge theories defined on an interval, see, for example,~\cite{Feverati:1998va,Rakovszky:2016ugs,Schmoll:2023eez,Houtz:2025lbv}. Extensions to higher spacetime dimensions have also been explored~\cite{Hogervorst2015,Elias-Miro:2020qwz}, as well as formulations in boosted frames~\cite{Chen:2022zms}.

Hamiltonian-truncation approaches have also been pursued in light-cone and conformal bases, where infinite-volume dynamics can be accessed efficiently within certain classes of models, see Refs.~\cite{Anand:2020gnn,Fitzpatrick:2022dwq}.
Across these various implementations, a common practical challenge is the rapid growth of the truncated basis with the energy cut-off \(E_{\rm max}\). This growth leads to an exponential increase in computational cost and complicates simple expectations for the scaling of truncation errors, thereby motivating the development of renormalized and improved variants of HT, in which the effects of high-energy states are systematically integrated out rather than simply discarded~\cite{Giokas:2011ix,Hogervorst2015,Rychkov2015,Rychkov2016,Elias_Mir_2016_RHT,Elias_Mir_2017_NLO,Elias_Mir_2017_HP}.

In parallel, alternative directions have been explored to mitigate the computational cost associated with large Hilbert spaces. One such direction is the use of Hamiltonian-truncation bases as qubit-efficient encodings for digital quantum simulations of real-time QFT dynamics and scattering processes~\cite{Ingoldby:2025bdb,Draper:2025nku}. 

Another complementary approach is provided by Hamiltonian truncation effective theory (HTET), introduced in Ref.~\cite{Cohen2022} and developed further in the subsequent works~\cite{Demiray:2025zqh,Delouche:2023wsl, EliasMiro:2022pua}, which treat the cut-off \(E_{\rm max}\) as an effective field theory (EFT) scale. States above \(E_{\rm max}\) are integrated out, yielding a finite-dimensional effective Hamiltonian acting on the truncated subspace. Corrections are organized as a controlled expansion in inverse powers of \(E_{\rm max}\), with an interplay between local terms and genuinely non-local contributions. 

In this work we further develop the HTET program along two main directions. First, building on Ref.~\cite{Cohen2022}, we develop an all-order resummation of the local HTET contributions, with the aim of improving the practical convergence of the effective Hamiltonian at fixed computational cost. Second, extending Ref.~\cite{Demiray:2025zqh}, we derive the second-order non-local corrections and reformulate the matching directly in the continuum, postponing space compactification. This reorganization makes the structure of the expansion more transparent and facilitates higher-order computations.

The paper is organized as follows. In Sec.~\ref{sec_intro_ht_htet} we review HT and its EFT reformulation, introducing the truncation via an energy cut-off \(E_{\rm max}\) and defining the truncated effective Hamiltonian. We then specialize to two-dimensional \(\lambda\phi^4\) theory as a testing model, detailing the construction of the Fock basis and the implementation of the truncation scheme. 

In Sec.~\ref{sec_resummation_of_local_expansion} we develop the resummation of local terms, 
by summing together infinite classes of diagrams that share a fixed topology.
The sums are performed to all orders and within the local approximation, yielding compact expressions for the matching corrections to the mass and quartic coupling. 
In Sec.~\ref{sec_NNLO_corrections} we derive sub-leading non-local contributions generated by the second-order matching, and formulate the matching directly in the continuum  to handle the associated distributional structures. Although these terms scale as \(\mathcal O(E_{\rm max}^{-4})\), they do not constitute the full matching at that order, which would also require local contributions from the next perturbative order.
Finally, in Sec.~\ref{sec_Numerical Results} we present a numerical analysis of the resulting effective Hamiltonians, assessing the impact of the resummed  and subleading non-local contributions on the convergence of low-energy spectral observables.

\section{Hamiltonian Truncation Effective Theory for the \texorpdfstring{\(\lambda \phi^4\)}{lambda phi4} theory in two dimensions}
\label{sec_intro_ht_htet}
In this section, we briefly introduce the HT method~\cite{Yurov1990,Yurov1991,Hogervorst2015}, and its formulation as HTET~\cite{Cohen2022}, in the case of the two-dimensional \(\lambda\phi^4\)  theory defined on a circle of circumference \(L = 2\pi R\). The basic idea is to map an infinite-dimensional problem into a finite-dimensional numerical one  by restricting the space of states to those with energies below an ultraviolet (UV) cut-off 
 \(E_{\max}\), while systematically controlling the truncation errors.

One starts by splitting the full Hamiltonian into a free, solvable part and an interaction term, 
\begin{equation}
\label{eq:HoandV}
H = H_0 + V,
\end{equation}
with \(H_0\) such that its spectrum and eigenstates are exactly known, i.e.,
\begin{equation}
H_0 \,|\mathcal{E}_i\rangle = \mathcal{E}_i\,|\mathcal{E}_i\rangle, \qquad \langle \mathcal{E}_i|\mathcal{E}_j\rangle = \delta_{ij}.
\end{equation}
The strategy is then to numerically diagonalize the full Hamiltonian \(H\) in the basis of eigenstates of \(H_0\), with the aim of extracting eigenvalues and eigenvectors defined by \(H\lvert \Psi \rangle = E \lvert \Psi \rangle\). This requires making the Hilbert space \(\mathcal{H}\), spanned by the free eigenstates, finite dimensional. To this end, one first compactifies the spatial dimensions so that the spectrum of \(H_0\) becomes discrete, and then imposes an energy cut-off \(E_{\rm max}\) on the free spectrum. This cut-off induces an orthogonal decomposition of \(\mathcal{H}\) into low-energy and high-energy subspaces, i.e., 
\begin{equation}
\mathcal{H}=\mathcal{H}_{\mathrm{tru}}\oplus \mathcal{H}_{\mathrm{neg}},
\qquad
\mathcal{H}_{\mathrm{tru}}=\mathrm{span}\{|\mathcal{E}_i\rangle\,:\,\mathcal{E}_i\le E_{\max}\},
\qquad
\mathcal{H}_{\mathrm{neg}}=\mathrm{span}\{|\mathcal{E}_i\rangle\,:\,\mathcal{E}_i>E_{\max}\}.
\end{equation}
By construction, \(\mathcal{H}_{\mathrm{tru}}\) is finite-dimensional and therefore suitable for numerical analysis.
It is then convenient to introduce the following projection operators
\begin{equation}
P \doteq \sum_{E_i\le E_{\max}} |\mathcal{E}_i\rangle\langle \mathcal{E}_i|, 
\qquad 
Q \doteq \sum_{E_i> E_{\max}} |\mathcal{E}_i\rangle\langle \mathcal{E}_i|,
\qquad P+Q=\mathds{1}.
\end{equation}
HT, hereafter referred to as \enquote{raw}, corresponds to simply approximating the full Hamiltonian as \(H \simeq PHP\).
The accuracy of this approximation improves as a power law in \(E_{\max}\). However, the computational cost required to achieve this improvement grows rapidly with the size of the basis, since the number of states included in \(\mathcal{H}_{\mathrm{tru}}\) scales exponentially with \(E_{\max}\).

Several methods have been developed to mitigate this problem by improving the convergence, see Refs.~\cite{Rychkov2015,Rychkov2016,Elias_Mir_2016_RHT,Elias_Mir_2017_NLO,Elias_Mir_2017_HP}.
HTET is one such approach, in which  an  EFT framework is employed to systematically improve the accuracy of the truncation. 
The key observation is that, for sufficiently large \(E_{\max}\), the UV dynamics associated with highly energetic states can be treated perturbatively, while the numerical diagonalization on \(\mathcal{H}_{\mathrm{tru}}\) captures the non-perturbative physics in the infrared (IR) sector.
This is precisely the setting in which the EFT viewpoint becomes natural, since the effects of the excluded states in \(\mathcal{H}_{\mathrm{neg}}\) can be absorbed into effective operators acting within the truncated theory.

HTET provides a systematic framework for this procedure by constructing an effective Hamiltonian \(H_{\mathrm{eff}} = H_0 + V_{\mathrm{eff}}\) defined on \(\mathcal{H}_{\mathrm{tru}}\), where
\begin{equation}
V_{\mathrm{eff}} = H_1 + H_2 + H_3 + \cdots, 
\qquad 
H_n = \mathcal{O}(V^n).
\end{equation}
The effective Hamiltonian \(H_{\mathrm{eff}}\) is obtained operationally by requiring that a low-energy observable agrees between the fundamental theory and the effective theory defined at the scale \(E_{\max}\).
The chosen observable is the transition matrix \(T\), defined from the time-evolution operator in the interaction picture and admitting a perturbative expansion in \(V\).
Let \(|i\rangle\) and \(|f\rangle\) denote low-energy external states, with \(\mathcal{E}_{i,f} \ll E_{\max}\), and introduce the energy differences
\begin{equation}
E_{fi} \doteq \mathcal{E}_f - \mathcal{E}_i,
\qquad 
E_{f\alpha} \doteq \mathcal{E}_f - \mathcal{E}_\alpha,
\end{equation}
where \(\mathcal{E}_\alpha\) is the energy of an intermediate state \(|\alpha\rangle\).
By imposing the equality between the fundamental and effective transition matrices order by order in \(V\), one obtains matching conditions for the operators \(H_n\).
At the lowest orders, these read
\begin{align}
\langle f|H_1|i\rangle_{\mathrm{eff}} &= \langle f|V|i\rangle, \\
\langle f|H_2|i\rangle_{\mathrm{eff}} &= \sum_{\alpha}^> \frac{\langle f|V|\alpha\rangle \langle \alpha|V|i\rangle}{\mathcal{E}_{f\alpha}},
\label{eq:H2_matching_intro}
\end{align}
where the notation \(\displaystyle \sum_{\alpha}^>\) indicates that the sum over intermediate states is restricted to those satisfying \(\mathcal{E}_\alpha > E_{\max}\).

To quantify the corrections to \(H_{\mathrm{eff}}\) from Eq.~\eqref{eq:H2_matching_intro},
we consider the two-dimensional \(\lambda\phi^4\) theory, defined by a quartic self-interaction with coupling constant \(\lambda\) and a mass parameter \(m\). Despite its simplicity, this model captures all the structural features relevant for HT.

In two dimensions, the scalar field has scaling dimension \([\phi]=0\), while the mass and the quartic coupling have mass dimensions \([m^2]=[\lambda]=2\). As a consequence, the interaction is strongly relevant, in the sense that its coupling has positive mass dimension and therefore grows rapidly toward the IR. This also implies that the UV behavior of the theory is particularly well controlled once normal ordering is imposed.

In finite volume, one works with the so-called \enquote{naive} Hamiltonian
\begin{equation}
H = H_0 + V
= \sum_{k}\omega_k^\pdagg\,a^\dagger_k a_k^\pdagg
+\frac{\lambda}{4!}\int_0^{L}dx\,:\phi^4(x):\;,
\qquad L=2\pi R.
\label{eq:H_naive}
\end{equation}
More precisely, one should account for Casimir-energy contributions and for terms arising from the mismatch between normal ordering defined at finite volume and in the infinite-volume limit. These effects are, however, exponentially suppressed in the regime \(Lm \gg 1\).

To discretize the spectrum of \(H_0\) and obtain a countable basis, the theory is defined on a cylinder \(S^1_R \times \mathbb{R}\) with periodic boundary conditions,
\(\phi(x+2\pi R,t)=\phi(x,t)\).
The compactification discretizes the momentum modes and allows for a Fourier expansion.
For a massive scalar field, one can write
\begin{equation}
\phi(x,t)=\frac{1}{\sqrt{2\pi R}}\sum_{k\in\mathbb{Z}}\frac{1}{\sqrt{2\omega_k}}
\left(a_k^\pdagg\,e^{-ik x/R}+a^\dagger_k\,e^{ik x/R}\right),
\qquad 
\omega_k=\sqrt{(k/R)^2+m^2},
\end{equation}
with \([a_k^\pdagg,a^\dagger_{k'}]=\delta_{kk'}\).
It is convenient to introduce positive and negative frequency components, corresponding to annihilation and creation operators:
\begin{equation}
\phi_k=\phi^{(+)}_k+\phi^{(-)}_{-k}, 
\qquad 
\phi^{(+)}_k=\frac{1}{\sqrt{2\omega_k}}\,a_k,
\qquad 
\phi^{(-)}_k=\frac{1}{\sqrt{2\omega_k}}\,a^\dagger_k.
\end{equation}
The eigenstates of \(H_0\) are Fock states, labeled by occupation numbers \(n_k\in\mathbb{N}_0\),
\begin{equation}
|\bar n\rangle \doteq |n_0,n_1,n_{-1},\dots\rangle
=\prod_{k\in\mathbb{Z}}\frac{(a^\dagger_k)^{n_k}}{\sqrt{n_k!}}\,|0\rangle,
\qquad 
H_0|\bar n\rangle=\Bigl(\sum_k \omega_k n_k\Bigr)|\bar n\rangle.
\end{equation}
Having established the truncated basis and the operator setup, we now turn to the explicit determination of the HTET matching corrections in the \(2d\) \(\lambda\phi^{4}\) theory. In the local approximation, the matching corrections can then be represented as a finite set of local operators (counterterms) acting within \(\mathcal{H}_{\mathrm{tru}}\), with coefficients that depend on \(E_{\max}\) but not on the external energies. Beyond this leading order, the matching acquires a residual dependence on 
\(\mathcal{E}_{i,f}\), and the corrections can no longer be represented purely in terms of local operators.
In this context, non-locality implies that the effective interaction becomes sensitive to the energies of the external states and therefore  to the presence of spectator particles that may not be spatially localized with the interacting system~\cite{Cohen2022}. 
These genuinely non-local contributions are, however, parametrically suppressed by additional powers of \(1/E_{\max}\).

\section{Resummation of local expansion}
\label{sec_resummation_of_local_expansion}
Reference~\cite{Cohen2022} carries out the matching procedure described in Sec.~\ref{sec_intro_ht_htet} explicitly up to order \(\mathcal{O}(V^2)\) using a diagrammatic approach. This analysis yields local corrections to the quartic coupling and to the mass, which significantly improve the convergence of Hamiltonian-truncation numerical calculations.

The purpose of this section is to extend this analysis by deriving resummed expressions for the matching corrections to both the quartic coupling and the mass, incorporating simultaneously the contributions from all orders. This is achieved by exploiting the structure of the diagrammatic expansion and by applying
standard resummation techniques to infinite classes of diagrams that share a fixed topology
and differ only in the number of internal loops. (For a general introduction to perturbative expansions and summation methods, see Ref.~\cite{ZinnJustin2019}, Chapter~23.)

We begin by focusing on a specific four external legs topology, which provides the most natural
starting point for an all-order generalization of the matching corrections to the effective
quartic coupling. Using the diagrammatic rules introduced in Ref.~\cite{Cohen2022}, the contribution of this diagram at generic order \(\mathcal{O}(V^n)\) can be written as
\begin{align}
\label{eq:diagram_D}
D^{(n)} \doteq \adjustbox{scale=0.4, valign=c}{
\begin{tikzpicture}[line width=1.5pt]

\def\s{1.5}

\begin{feynman}
    \vertex (b);
    \vertex [above left=\s cm of b] (a3) {$\scalemath{2.5}{3}$};
    \vertex [below left=\s cm of b] (a4) {$\scalemath{2.5}{4}$};
    \vertex (c) [right=\s*1.25 cm of b];
    \vertex (d) [right=\s*1.25 cm of c];
    \vertex (e) [right=\s*1.25 cm of d];
    \vertex (f) [right=\s*1.25 cm of e];
    \vertex [below right=\s cm of f] (a2) {$\scalemath{2.5}{2}$};
    \vertex [above right=\s cm of f] (a1) {$\scalemath{2.5}{1}$};

    \adjustbox{scale=0.8, valign=c}{\includestandalone{Diagrams/*}} {
        (a3) -- (b),
        (a4) -- (b),
        (f) -- (a1),
        (f) -- (a2),
    };

    \pgfmathsetmacro{\rad}{0.625*\s} 
    \draw (e) ++(0,0) arc[start angle=180, end angle=0, 
        x radius=\rad cm, y radius=\rad cm]
        node[midway, above] {$\scalemath{2.5}{5}$};
    \draw (e) ++(0,0) arc[start angle=180, end angle=0, 
        x radius=\rad cm, y radius=-\rad cm]
        node[midway, below] {$\scalemath{2.5}{6}$};
    \draw (d) ++(0,0) arc[start angle=180, end angle=0, 
        x radius=\rad cm, y radius=\rad cm]
        node[midway, above] {$\scalemath{2.5}{7}$};
    \draw (d) ++(0,0) arc[start angle=180, end angle=0, 
        x radius=\rad cm, y radius=-\rad cm]
        node[midway, below] {$\scalemath{2.5}{8}$};
    \draw (b) ++(0,0) arc[start angle=180, end angle=0, 
        x radius=\rad cm, y radius=\rad cm]
        node[midway, above] {$\scalemath{2.5}{2n+1}$};
    \draw (b) ++(0,0) arc[start angle=180, end angle=0, 
        x radius=\rad cm, y radius=-\rad cm]
        node[midway, below] {$\scalemath{2.5}{2n+2}$};

    \draw[fill=black] (b) circle (\s*1.5 pt);
     \draw[fill=black] (d) circle (\s*1.5 pt);
    \draw[fill=black] (c) circle (\s*1.5 pt);
    \draw[fill=black] (f) circle (\s*1.5 pt);
    \draw[fill=black] (e) circle (\s*1.5 pt);
\node at ($(c)!0.5!(d)$) {$\scalemath{2.5}{\cdots}$};


\end{feynman}

\end{tikzpicture}
}
&=
\frac{1}{2}\left(\frac{\lambda}{4\pi R}\right)^n
\sum_{1\ldots 2(n+1)}
\delta_{12,34}\,
\bra{f} \phi^-_4 \phi^-_3 \phi^+_2 \phi^+_1 \ket{i}
\notag \\
&\quad\times
\frac{
\displaystyle
\prod_{j=0}^{n-2}
\delta_{34,(5+2j)(6+2j)}
}{
\displaystyle
\left(\prod_{j=5}^{2(n+1)} 2\omega_j\right)
\left(\prod_{j=0}^{n-2}
(\omega_3+\omega_4-\omega_{5+2j}-\omega_{6+2j})\right)
}\, .
\end{align}
Here we have introduced a shorthand notation for the momenta, \(k_i = i\), and for the Kronecker delta, \(\delta_{k_1 + k_2 = k_3 + k_4} = \delta_{12,34}\).
In the effective theory, the same diagram is evaluated under the constraint that all
intermediate states have energies below the cut-off \(E_{\rm max}\). At order
\(\mathcal{O}(V^n)\), the result is therefore multiplied by a product of Heaviside
step functions which enforce these energy restrictions:
\begin{equation}
\label{eq:theta_diagram_Deff}
D^{(n)}\vert_{\rm eff}
\;\propto\;
\prod_{j=1}^{n-1}
\Theta\!\left(
E_{\rm max}
-
\mathcal{E}_f
+
\omega_3+\omega_4
-
\omega_{2j+3}
-
\omega_{2j+4}
\right).
\end{equation}
Following Ref.~\cite{Cohen2022}, we simplify the non-trivial dependence on the external energies by working in the local approximation. Since we are interested in low-energy initial and final states,
\(\mathcal{E}_{i,f} \ll E_{\rm max}\), the dependence on external energies and momenta can be systematically expanded. Retaining only the leading term in this expansion amounts to evaluating the diagram at vanishing external energies and momenta. Within this approximation, the contribution of the diagram reduces to
\begin{align}
\label{eq:diagram_D_local}
D^{(n)}
&\simeq
-
\left(-\frac{\lambda}{4\pi R}\right)^n
\pi R
\int \mathrm{d}x\;
\bra{f}
\left[\phi^-\right]^2
\left[\phi^+\right]^2
\ket{i}
\sum_{k_1,k_2,\dots,k_{n-1}}
\frac{1}{
\displaystyle
\prod_{j=1}^{n-1}
(2\omega_{k_j})^3
}\, .
\end{align}
It is convenient to introduce the number of internal loops as an index, defined as
\(L \doteq n-1\).  In the local approximation, the \(L\)-loop contribution factorizes into \(L\) identical momentum sums, so that the all-order sum reduces to a geometric series:
\begin{align}
\label{eq:series_diagram_D}
\sum_{L=1}^{\infty} D^{(L+1)}
&\simeq
-\pi R \int \mathrm{d}x\;
\bra{f}\left[\phi^-\right]^2\left[\phi^+\right]^2\ket{i}\,
\sum_{L=1}^{\infty}\left(-\frac{\lambda}{4\pi R}\right)^{L+1}
\sum_{k_1,\dots,k_L}\prod_{j=1}^{L}\frac{1}{(2\omega_{k_j})^3}
\notag\\
&=
\frac{\lambda}{4}\int \mathrm{d}x\;
\bra{f}\left[\phi^-\right]^2\left[\phi^+\right]^2\ket{i}\,
\sum_{L=1}^{\infty}
\left(-\frac{\lambda}{4\pi R}\sum_{k}\frac{1}{(2\omega_k)^3}\right)^{L}
\notag\\
&=
\frac{\lambda}{4}\int \mathrm{d}x\;
\bra{f}\left[\phi^-\right]^2\left[\phi^+\right]^2\ket{i}\;
\frac{X}{1-X},
\end{align}
where we defined
\begin{equation}
\label{eq:def_X}
X \doteq -\frac{\lambda}{4\pi R}\sum_{k}\frac{1}{(2\omega_k)^3}\, .
\end{equation}
The effective-theory result is obtained analogously, with the only modification coming from
the cut-off constraint in Eq.~\eqref{eq:theta_diagram_Deff}, which amounts to replacing \(X\)
by
\begin{equation}
\label{eq:def_Xeff}
X^{\rm (eff)}\doteq -\frac{\lambda}{4\pi R}\sum_{k}\frac{\Theta(E_{\rm max}-2\omega_k)}{(2\omega_k)^3}\, .
\end{equation}
Combining the full and effective expressions, we obtain
\begin{align}
\label{eq:sirres_D-Deff}
\sum_{L=1}^{\infty} D^{(L+1)}-\sum_{L=1}^{\infty} D^{(L+1)}\vert_{\rm eff}
&\simeq
\frac{\lambda}{4}\int \mathrm{d}x\;
\bra{f}\left[\phi^-\right]^2\left[\phi^+\right]^2\ket{i}\;
\left[\frac{X-X^{\rm (eff)}}{(1-X)(1-X^{\rm (eff)})}\right].
\end{align}
This immediately yields the following result for the resummed matching correction to the quartic coupling
\begin{equation}
\label{eq:delta_lambda_resum}
\delta\tilde{\lambda}
=
6\lambda\left[\frac{X-X^{\rm (eff)}}{(1-X)(1-X^{\rm (eff)})}\right].
\end{equation}

We now turn to a two external legs topology, which yields the local matching correction associated
to the mass.
Generalizing this diagram to order \(\mathcal{O}(V^n)\), one finds
\begingroup
\setlength{\abovedisplayskip}{-10pt}
\begin{align}
\label{eq:diagram_B}
B^{(n)} \doteq \adjustbox{scale=0.4, valign=c}{
\begin{tikzpicture}[line width=1.5pt]

\def\s{1.5}

\begin{feynman}
    \vertex (a) {$\scalemath{2.5}{2}$};
    \vertex (b) [right=\s*1.25 cm of a];
    \vertex (l1) [right=\s*1.25 cm of b];  
    \vertex (l2) [right=\s*1.25 cm of l1];
    \vertex (l3) [right=\s*1.25 cm of l2];
    \vertex (c) [right=\s*1.25 cm of l3];
    \vertex (d) [right=\s*1.25 cm of c]{$\scalemath{2.5}{1}$};

\pgfmathsetmacro{\bbheight}{4.5*\s} 
\pgfmathsetmacro{\bbhalf}{0.5*\bbheight}
\path
  ($(b)+(-0.5cm,\bbhalf cm)$) rectangle
  ($(c)+( 0.5cm,-\bbhalf cm)$);

    \adjustbox{scale=0.8, valign=c}{\includestandalone{Diagrams/*}} {
        (a) -- (b),
        (c) -- (d),

    };

    \pgfmathsetmacro{\rad}{0.625*\s} 
    \draw (b) ++(0,0) arc[start angle=180, end angle=0, 
        x radius=\rad cm, y radius=\rad cm]
        node[midway, above] {$\scalemath{2.5}{2n}$};
    \draw (b) ++(0,0) arc[start angle=180, end angle=0, 
        x radius=\rad cm, y radius=-\rad cm]
         node[midway, below, xshift=+35pt] {$\scalemath{2.5}{2n+1}$};
    \draw (l3) ++(0,0) arc[start angle=180, end angle=0, 
        x radius=\rad cm, y radius=\rad cm]
        node[midway, above] {$\scalemath{2.5}{6}$};
    \draw (l3) ++(0,0) arc[start angle=180, end angle=0, 
        x radius=\rad cm, y radius=-\rad cm]
        node[midway, below] {$\scalemath{2.5}{7}$};
    \draw (l2) ++(0,0) arc[start angle=180, end angle=0, 
        x radius=\rad cm, y radius=\rad cm]
        node[midway, above] {$\scalemath{2.5}{4}$};
    \draw (l2) ++(0,0) arc[start angle=180, end angle=0, 
        x radius=\rad cm, y radius=-\rad cm]
        node[midway, below] {$\scalemath{2.5}{5}$};

    \pgfmathsetmacro{\radU}{0.625*\s*4} 
    \pgfmathsetmacro{\depth}{1.75*\s}  
    \draw (b) arc[start angle=180, end angle=0,
        x radius=\radU cm, y radius=-\depth cm] 
        node[midway, below] {$\scalemath{2.5}{3}$};;
        
    \draw[fill=black] (b) circle (\s*1.5 pt);
    \draw[fill=black] (c) circle (\s*1.5 pt);
    \draw[fill=black] (l1) circle (\s*1.5 pt);
    \draw[fill=black] (l2) circle (\s*1.5 pt);
    \draw[fill=black] (l3) circle (\s*1.5 pt);

\node at ($(l1)!0.5!(l2)$) {$\scalemath{2.5}{\cdots}$};

\end{feynman}

\end{tikzpicture}
}
&=
\frac{\lambda}{18\pi R}\left(\frac{3\lambda}{4\pi R}\right)^{n-1}
\sum_{1\ldots 2n+1}\delta_{1,2}\,
\bra{f}\phi^-_2\phi^+_1\ket{i}
\notag\\[-6pt]
&\quad\times
\frac{1}{2\omega_3}
\prod_{j=1}^{n-1}
\frac{
\displaystyle \delta_{1,(2+2j)(3+2j)(3)}
}{
\displaystyle
2\omega_{2+2j}\;2\omega_{3+2j}\,
\bigl(\omega_2-\omega_3-\omega_{2+2j}-\omega_{3+2j}\bigr)
}\, .
\end{align}
As before, in the effective theory the restriction to intermediate states with total energy
below the cut-off is implemented through a product of step functions multiplying  the entire expression, i.e., 
\endgroup
\begin{align}
\label{eq:theta_diagram_Beff}
B^{(n)}\vert_{\rm eff}
\;\propto\;
\prod_{j=1}^{n-1}
\Theta\!\left(
E_{\rm max}
-\mathcal{E}_f
+\omega_2-\omega_3-\omega_{2+2j}-\omega_{3+2j}
\right).
\end{align}
Working again in the local approximation, the
all-order sum takes the form
\begin{align}
\label{eq:series_diagram_B}
\sum_{L=1}^{\infty} B^{(L+1)}
&\simeq
\frac{\lambda}{18\pi R}\int \mathrm{d}x\;
\bra{f}\phi^-\phi^+\ket{i}\;
\sum_{L=1}^{\infty}\left(\frac{3\lambda}{4\pi R}\right)^{L}
\sum_k \frac{1}{2\omega_k}\,\bigl(G_k\bigr)^{L}
\notag\\
&=
\frac{\lambda}{18\pi R}\int \mathrm{d}x\;
\bra{f}\phi^-\phi^+\ket{i}\;
\sum_k \frac{1}{2\omega_k}\,\frac{N G_k}{1-NG_k},
\end{align}
where we defined
\begin{equation}
\label{eq:def_Gk_N}
N\doteq \frac{3\lambda}{4\pi R}
\quad \text{and}\quad
G_k \doteq
-\sum_j
\frac{1}{2\omega_j\,2\omega_{k+j}\,(\omega_k+\omega_j+\omega_{k+j})}\, .
\end{equation}
The effective-theory result is obtained by enforcing the cut-off constraint from
Eq.~\eqref{eq:theta_diagram_Beff} directly in the loop sum, namely by replacing
\(G_k \to G_k^{\rm (eff)}\)with
\begin{equation}
\label{eq:def_Gkeff}
G_k^{\rm (eff)} \doteq
-\sum_j
\frac{\Theta(E_{\rm max}-\omega_k-\omega_j-\omega_{k+j})}
{2\omega_j\,2\omega_{k+j}\,(\omega_k+\omega_j+\omega_{k+j})}\, .
\end{equation}
Combining the contributions of the fundamental and effective theories, one obtains
\begin{align}
\sum_{L=1}^{\infty} B^{(L+1)}-\sum_{L=1}^{\infty} B^{(L+1)}\vert_{\rm eff}
&=
\frac{\lambda}{18\pi R}\int \mathrm{d}x\;
\bra{f}\phi^-\phi^+\ket{i}\;
\sum_k \frac{N}{2\omega_k}
\left[
\frac{G_k}{1-NG_k}
-
\frac{G_k^{\rm (eff)}}{1-NG_k^{\rm (eff)}}
\right].
\end{align}
This yields the resummed matching correction to the mass parameter,
\begin{equation}
\label{eq:delta_m_resum}
\delta\tilde{m}^2
=
\frac{\lambda}{9\pi R}\sum_k \frac{N}{2\omega_k}
\left[
\frac{G_k}{1-NG_k}
-
\frac{G_k^{\rm (eff)}}{1-NG_k^{\rm (eff)}}
\right].
\end{equation}
In conclusion, the resummed matching corrections to the quartic coupling and to the mass, given in Eqs.~\eqref{eq:delta_lambda_resum} and \eqref{eq:delta_m_resum}, respectively, provide compact all-order expressions that capture the dominant high-energy effects and will be implemented numerically in Sec.~\ref{sec_Numerical Results} to assess their quantitative impact on Hamiltonian-truncation results.

\section{Next-to-next-to-local corrections}
\label{sec_NNLO_corrections}
Building on the framework of Ref.~\cite{Demiray:2025zqh}, subleading 
corrections arising from the effective 
matching procedure can be systematically 
derived. In particular, non-local 
contributions at order  $\mathcal{O}(V^2)$  were shown to scale as
$\mathcal{O}(E_{\rm max}^{-3})$, 
 providing a further improvement over the 
 leading local corrections, which already 
 enhance the convergence of the bare 
 HT. The resulting 
 framework naturally organizes the 
 corrections into a systematic expansion 
 in inverse powers of the truncation 
 scale $E_{\rm max}$.

The expansion is organized as follows: the leading local contribution at order $\mathcal O(V^n)$ starts at $\mathcal O(E_{\max}^{-(2n-2)})$, whereas the first genuinely non-local contribution at the same perturbative order is further suppressed and starts at $\mathcal O(E_{\max}^{-(2n-1)})$. For clarity, Table~\ref{tab:htet-bookkeeping} summarizes the structure of the expansion up to $\mathcal O(E_{\max}^{-6})$. It follows that $\mathcal O(E_{\max}^{-4})$ receives two distinct types of contributions: the second non-local correction in the $\mathcal O(V^2)$ sector and the leading local correction in the $\mathcal O(V^3)$ sector. In the present work we compute only the former, which we refer to hereafter as the next-to-next-to-local corrections (NNLoc). Therefore, we are not performing the full matching at $\mathcal O(E_{\max}^{-4})$, since the local $\mathcal O(V^3)$ contributions are not included. These  terms are expected to be standard local HTET counterterms of order $\lambda^3/E_{\max}^4$, built from the local operator basis allowed by the symmetries~\cite{Cohen2022}. While their derivation is conceptually straightforward, following from the matching condition, it is technically cumbersome due to the large number of third-order diagrams involved; we therefore leave their evaluation for future work.

\begin{table}[ht]
\centering
\renewcommand{\arraystretch}{1.25}
\begin{tabular}{c|c|c|c}
\hline
\(1/E_{\max}\) 
& \(\mathcal O(V^2)\) 
& \(\mathcal O(V^3)\) 
& \(\mathcal O(V^4)\) \\
\hline
\(\mathcal O(E_{\max}^{-2})\) 
& leading local 
& 
& 
\\

\(\mathcal O(E_{\max}^{-3})\) 
& first non-local 
& 
& 
\\

\(\mathcal O(E_{\max}^{-4})\) 
& second non-local
& leading local 
& 
\\

\(\mathcal O(E_{\max}^{-5})\) 
& third non-local 
& first non-local 
& 
\\

\(\mathcal O(E_{\max}^{-6})\) 
& fourth non-local 
& second non-local 
& leading local 
\\
\hline
\end{tabular}
\caption{Schematic organization of the HTET expansion. Rows indicate the order in \(1/E_{\max}\), while columns indicate the perturbative order in \(V\). The contribution computed in the present work is the second non-local term in the \(\mathcal O(V^2)\) sector, which contributes at \(\mathcal O(E_{\max}^{-4})\).}
\label{tab:htet-bookkeeping}
\end{table}

We begin by classifying all diagrams with four external legs into nine distinct topologies, which differ in the construction of the intermediate state: 
\begin{equation}
\label{diagram:all4legs_abcd}
\begin{aligned}
 \adjustbox{scale=0.4, valign=c}{\begin{tikzpicture}[line width=1.5pt]

\def\s{1.5}
\begin{feynman}
    \vertex (b);
    \vertex [above left=\s cm of b] (a3) {$\scalemath{2.5}{a}$};
    \vertex [below left=\s cm of b] (a4){$\scalemath{2.5}{b}$};
    \vertex (c)[right=\s*1.25 cm of b];
    \vertex [below right=\s cm of c] (a2) {$\scalemath{2.5}{d}$};
    \vertex [above right=\s cm of c] (a1){$\scalemath{2.5}{c}$};

    \adjustbox{scale=0.8, valign=c}{\includestandalone{Diagrams/*}} {
        (a3) -- (b),
        (a4) -- (b),
        (b) -- [half left,looseness=1,edge label=$\scalemath{2.5}{}$] (c) -- [half left,looseness=1,edge label=$\scalemath{2.5}{}$](b),
        (c) -- (a1),
        (c) -- (a2),
    };
    \draw[fill=black] (b) circle (\s*1.5 pt);
    \draw[fill=black] (c) circle (\s*1.5 pt);
\end{feynman}
\end{tikzpicture}} \quad\quad
 \adjustbox{scale=0.4, valign=c}{\begin{tikzpicture}[line width=1.5pt]

\def\s{1.5}
\def\t{90}
\def\d{0}

\begin{feynman}
    \vertex (b);
    \vertex [above left=\s cm of b] (a3) {$\scalemath{2.5}{a}$};
    \vertex [below left=\s cm of b] (a4) {$\scalemath{2.5}{b}$};

    \vertex (c) [right={1.25*\s cm} of b];

    \vertex [above right=\s cm of c] (a1) {$\scalemath{2.5}{c}$};
    \vertex [below right=\s cm of c] (a2) {$\scalemath{2.5}{d}$};

    \adjustbox{scale=0.8, valign=c}{\includestandalone{Diagrams/*}} {
        (a3) -- [in=\t, out=\d] (c),
        (a4) -- [in=-\t, out=-\d] (c),

        (b)  -- [half left, looseness=1] (c)
             -- [half left, looseness=1] (b),

        (a1) -- [in=\t, out=180] (b),
        (a2) -- [in=-\t, out=-180] (b),
    };

    \draw[fill=black] (b) circle ({1.5*\s pt});
    \draw[fill=black] (c) circle ({1.5*\s pt});
\end{feynman}

\end{tikzpicture}} \quad\quad
 \adjustbox{scale=0.4, valign=c}{\begin{tikzpicture}[line width=1.5pt]

\def\s{1.5}
\def\t{90}
\def\d{0}

\begin{feynman}
    \vertex (b);
    \vertex [below left=\s cm of b] (a3) {$\scalemath{2.5}{b}$};
    \vertex [ left=\s*0.70710678118 cm of b] (a4) {$\scalemath{2.5}{a}$};

    \vertex (c) [right={1.25*\s cm} of b];

    \vertex [above right=\s cm of c] (a2) {$\scalemath{2.5}{c}$};
    \vertex [right=\s*0.70710678118 cm of c] (a1) {$\scalemath{2.5}{d}$};

    \adjustbox{scale=0.8, valign=c}{\includestandalone{Diagrams/*}} {
        (a3) -- [in=-\t, out=-\d] (c),
        (a4) -- (b),

        (b)  -- [half left, looseness=1] (c)
             -- [half left, looseness=1] (b),

        (a1) -- (c),
        (a2) -- [in=\t, out=-180] (b),
    };

    \draw[fill=black] (b) circle ({1.5*\s pt});
    \draw[fill=black] (c) circle ({1.5*\s pt});
\end{feynman}

\end{tikzpicture}} \quad\quad  
 \adjustbox{scale=0.4, valign=c}{\begin{tikzpicture}[line width=1.5pt]

\def\s{1.5}
\def\t{90}
\def\d{0}

\begin{feynman}
    \vertex (b);

\def\ang{25}

\vertex (a3) at ($(b)+(-{1.3*\s*cos(\ang)}, {1.3*\s*sin(\ang)})$) {$\scalemath{2.5}{b}$};
\vertex (a4) at ($(b)+(-{1.3*\s*cos(\ang)}, {-1.3*\s*sin(\ang)})$) {$\scalemath{2.5}{c}$};

    \vertex (c) [right={1.25*\s cm} of b];

    \vertex [below=\s cm of b] (aux);
    \vertex [left=\s cm of aux] (a2) {$\scalemath{2.5}{d}$};

    \vertex [above=\s cm of b] (aux2);
    \vertex [left=\s cm of aux2] (a1) {$\scalemath{2.5}{a}$};

    \adjustbox{scale=0.8, valign=c}{\includestandalone{Diagrams/*}} {
        (a3) -- (b),
        (a4) -- (b),

        (b) -- [half left, looseness=1] (c)
             -- [half left, looseness=1] (b),

        (a1) -- [in=\t, out=\d] (c),
        (a2) -- [in=-\t, out=0] (c),
    };

    \draw[fill=black] (b) circle ({1.5*\s pt});
    \draw[fill=black] (c) circle ({1.5*\s pt});
\end{feynman}

\end{tikzpicture}} \quad\quad  
 \adjustbox{scale=0.4, valign=c}{\begin{tikzpicture}[line width=1.5pt]

\def\s{1.5}
\def\t{90}
\def\d{0}

\begin{feynman}
    \vertex (b);
    \vertex (c) [right={1.25*\s cm} of b];

    \def\ang{25}

    \vertex (a1) at ($(c)+({1.3*\s*cos(\ang)}, {1.3*\s*sin(\ang)})$) {$\scalemath{2.5}{b}$};
    \vertex (a2) at ($(c)+({1.3*\s*cos(\ang)}, {-1.3*\s*sin(\ang)})$) {$\scalemath{2.5}{c}$};

    \vertex [above=\s cm of c] (aux1);
    \vertex [right=\s cm of aux1] (a3) {$\scalemath{2.5}{a}$};

    \vertex [below=\s cm of c] (aux2);
    \vertex [right=\s cm of aux2] (a4) {$\scalemath{2.5}{d}$};

    \adjustbox{scale=0.8, valign=c}{\includestandalone{Diagrams/*}} {
        (b) -- [half left, looseness=1] (c)
            -- [half left, looseness=1] (b),

        (c) -- (a1),
        (c) -- (a2),
        (b) -- [in=\t+90, out=90] (a3),
        (b) -- [in=-\t-90, out=-90] (a4),
    };

    \draw[fill=black] (b) circle ({1.5*\s pt});
    \draw[fill=black] (c) circle ({1.5*\s pt});
\end{feynman}

\end{tikzpicture}} \\
 \adjustbox{scale=0.4, valign=c}{\begin{tikzpicture}[line width=1.5pt]

\def\s{1.5}
\def\t{90}
\def\d{0}
\def\ang{25}

\begin{feynman}
    \vertex (b);

\def\ang{25}

 \vertex [left=\s cm of b] (a4){$\scalemath{2.5}{b}$};
    \vertex (c) [right={1.25*\s cm} of b];

    \vertex [below=\s cm of b] (aux);
    \vertex [left=\s cm of aux] (a2) {$\scalemath{2.5}{c}$};

    \vertex [above=\s cm of b] (aux2);
    \vertex [left=\s cm of aux2] (a1) {$\scalemath{2.5}{a}$};

    \vertex [above=\s cm of c] (aux3);
    \vertex [right=\s cm of aux3] (a3) {$\scalemath{2.5}{d}$};

    \adjustbox{scale=0.8, valign=c}{\includestandalone{Diagrams/*}} {
        (a3) -- [in=\t, out=180](b),
        (a4) -- (b),

        (b) -- [half left, looseness=1] (c)
             -- [half left, looseness=1] (b),

        (a1) --[in=\t, out=0] (c),
        (a2) -- [in=-\t, out=0] (c),
    };

    \draw[fill=black] (b) circle ({1.5*\s pt});
    \draw[fill=black] (c) circle ({1.5*\s pt});
\end{feynman}

\end{tikzpicture}} \quad\quad 
 \adjustbox{scale=0.4, valign=c}{\begin{tikzpicture}[line width=1.5pt]

\def\s{1.5}
\def\t{90}
\def\d{0}
\def\ang{25}

\begin{feynman}
    \vertex (b);

\def\ang{25}

    \vertex (c) [right={1.25*\s cm} of b];

    \vertex [above=\s cm of b] (aux2);
    \vertex [left=\s cm of aux2] (a1) {$\scalemath{2.5}{a}$};

    \vertex [above=\s cm of c] (aux3);
    \vertex [right=\s cm of aux3] (a3) {$\scalemath{2.5}{b}$};
     \vertex [right=\s cm of c] (a4){$\scalemath{2.5}{c}$};

    \vertex [below=\s cm of c] (aux);
    \vertex [right=\s cm of aux] (a2) {$\scalemath{2.5}{d}$};

    \adjustbox{scale=0.8, valign=c}{\includestandalone{Diagrams/*}} {
        (a3) -- [in=\t, out=180](b),
        (a4) -- (c),

        (b) -- [half left, looseness=1] (c)
             -- [half left, looseness=1] (b),

        (a1) --[in=\t, out=0] (c),
        (a2) -- [in=-\t, out=180] (b),
    };

    \draw[fill=black] (b) circle ({1.5*\s pt});
    \draw[fill=black] (c) circle ({1.5*\s pt});
\end{feynman}

\end{tikzpicture}} \quad\quad
 \adjustbox{scale=0.4, valign=c}{\begin{tikzpicture}[line width=1.5pt]

\def\s{1.5}
\def\t{90}
\def\d{0}

\begin{feynman}

\vertex (b);

\def\ang{25}

\vertex (a3) at ($(b)+(-{1.3*\s*cos(\ang)}, {1.3*\s*sin(\ang)})$)
  {$\scalemath{2.5}{b}$};

\vertex (a4) at ($(b)+(-{1.3*\s*cos(\ang)}, {-1.3*\s*sin(\ang)})$)
  {$\scalemath{2.5}{c}$};

\vertex (c) [right={1.25*\s cm} of b];

\vertex [right=\s cm of c] (a2) {$\scalemath{2.5}{d}$};

\vertex [above=\s cm of b] (aux2);
\vertex [left=\s cm of aux2] (a1) {$\scalemath{2.5}{a}$};

\vertex [below=\s cm of b] (aux3);
\vertex [left=\s cm of aux3] (aux4) {$\scalemath{2.5}{\phantom{x}}$};

\adjustbox{scale=0.8, valign=c}{\includestandalone{Diagrams/*}}{
  (a3) -- (b),
  (a4) -- (b),
  (b) -- [half left, looseness=1] (c)
      -- [half left, looseness=1] (b),
  (a1) -- [in=\t, out=\d] (c),
  (a2) --(c),
};

\draw[fill=black] (b) circle ({1.5*\s pt});
\draw[fill=black] (c) circle ({1.5*\s pt});

\end{feynman}
\end{tikzpicture}} \quad\quad 
 \adjustbox{scale=0.4, valign=c}{\begin{tikzpicture}[line width=1.5pt]

\def\s{1.5}
\def\t{90}
\def\d{0}

\begin{feynman}
    \vertex (b);

    \vertex [left=\s cm of b] (a1) {$\scalemath{2.5}{a}$};

    \vertex (c) [right={1.25*\s cm} of b];

    \def\ang{25}
    \vertex (a3) at ($(c)+({1.3*\s*cos(\ang)}, {1.3*\s*sin(\ang)})$) {$\scalemath{2.5}{b}$};
    \vertex (a4) at ($(c)+({1.3*\s*cos(\ang)}, {-1.3*\s*sin(\ang)})$){$\scalemath{2.5}{c}$};

    \vertex [below=\s cm of c] (aux);
    \vertex [right=\s cm of aux] (a2) {$\scalemath{2.5}{d}$};
    \vertex [above=\s cm of c] (auxx);
\vertex [right=\s cm of auxx] (auxy) {$\scalemath{2.5}{\phantom{x}}$};
    \adjustbox{scale=0.8, valign=c}{\includestandalone{Diagrams/*}} {
        (a1) -- (b),

        (b) -- [half left, looseness=1] (c)
             -- [half left, looseness=1] (b),

        (a3) -- (c),
        (a4) -- (c),
        (a2) -- [in=-\t, out=180] (b),
    };

    \draw[fill=black] (b) circle ({1.5*\s pt});
    \draw[fill=black] (c) circle ({1.5*\s pt});
\end{feynman}

\end{tikzpicture}}
\end{aligned}
\end{equation}

To extract the NNLoc corrections, it is necessary to account for all possible symmetrizations of these diagrams, which are understood as inequivalent assignments of momentum indices for the intermediate states.

For simplicity, we focus on a single case with a four-leg external state, which admits only one topology.
In this case, one must consider all six inequivalent pairs \((b,c)\), defined as
\begin{equation}
(b,c)\in \mathcal{C}\doteq \{(1,2),(1,3),(1,4),(2,3),(2,4),(3,4)\}.
\end{equation}
Following the diagrammatic rules of Ref.~\cite{Cohen2022}, we obtain
 \begin{align}
    \label{eq:40legs}
     \adjustbox{scale=0.4, valign=c}{\begin{tikzpicture}[line width=1.5pt]

\def\s{1.5}
\def\t{90}
\def\d{0}

\begin{feynman}
    \vertex (b);

\def\ang{25}

\vertex (a3) at ($(b)+(-{1.3*\s*cos(\ang)}, {1.3*\s*sin(\ang)})$) {$\scalemath{2.5}{b}$};
\vertex (a4) at ($(b)+(-{1.3*\s*cos(\ang)}, {-1.3*\s*sin(\ang)})$) {$\scalemath{2.5}{c}$};

    \vertex (c) [right={1.25*\s cm} of b];

    \vertex [below=\s cm of b] (aux);
    \vertex [left=\s cm of aux] (a2) {$\scalemath{2.5}{d}$};

    \vertex [above=\s cm of b] (aux2);
    \vertex [left=\s cm of aux2] (a1) {$\scalemath{2.5}{a}$};

    \adjustbox{scale=0.8, valign=c}{\includestandalone{Diagrams/*}} {
        (a3) -- (b),
        (a4) -- (b),

        (b) -- [half left, looseness=1] (c)
             -- [half left, looseness=1] (b),

        (a1) -- [in=\t, out=\d] (c),
        (a2) -- [in=-\t, out=0] (c),
    };

    \draw[fill=black] (b) circle ({1.5*\s pt});
    \draw[fill=black] (c) circle ({1.5*\s pt});
\end{feynman}

\end{tikzpicture}} - \left[ \adjustbox{scale=0.4, valign=c}{\begin{tikzpicture}[line width=1.5pt]

\def\s{1.5}
\def\t{90}
\def\d{0}

\begin{feynman}
    \vertex (b);

\def\ang{25}

\vertex (a3) at ($(b)+(-{1.3*\s*cos(\ang)}, {1.3*\s*sin(\ang)})$) {$\scalemath{2.5}{b}$};
\vertex (a4) at ($(b)+(-{1.3*\s*cos(\ang)}, {-1.3*\s*sin(\ang)})$) {$\scalemath{2.5}{c}$};

    \vertex (c) [right={1.25*\s cm} of b];

    \vertex [below=\s cm of b] (aux);
    \vertex [left=\s cm of aux] (a2) {$\scalemath{2.5}{d}$};

    \vertex [above=\s cm of b] (aux2);
    \vertex [left=\s cm of aux2] (a1) {$\scalemath{2.5}{a}$};

    \adjustbox{scale=0.8, valign=c}{\includestandalone{Diagrams/*}} {
        (a3) -- (b),
        (a4) -- (b),

        (b) -- [half left, looseness=1] (c)
             -- [half left, looseness=1] (b),

        (a1) -- [in=\t, out=\d] (c),
        (a2) -- [in=-\t, out=0] (c),
    };

    \draw[fill=black] (b) circle ({1.5*\s pt});
    \draw[fill=black] (c) circle ({1.5*\s pt});
\end{feynman}

\end{tikzpicture}}\right]_{\rm eff} = \frac{1}{8} \left(\frac{\lambda}{2\pi R}\right)^2 \frac{1}{\dim(\mathcal{C})}\sum_{\substack{a,d,5,6 \\ (b,c)\in\mathcal{C}}}& \delta_{abcd,0}\;\delta_{bc,56}\;\bra{f} \phi_{d}^{(-)} \phi_{c}^{(-)}  \phi_{b}^{(-)} \phi_{a}^{(-)}\; \ket{i}\notag\\&\times\frac{\Theta(\mathcal{E}_f - \omega_b-\omega_c+\omega_5+\omega_6 - E_{\rm max})}{2\omega_5 2 \omega_6\,(\omega_b+\omega_c-\omega_5-\omega_6)},
\end{align}
where \(a,d\in\{1,2,3,4\}\), and with all indices mutually distinct, i.e.,
\(a\neq b\neq c\neq d\).
For other topologies, one obtains expressions that have similar structure to Eq.~\eqref{eq:40legs}. The results
can be systematically expanded by treating the IR scales
\(\omega_{a,b,c,d}\) and the external energies \(\mathcal{E}_{f,i}\) as small compared to
the UV scale set by the cut-off.  Moreover, the momentum-conservation constraint \(\delta_{bc,56}\) enforces
\(\omega_5\simeq\omega_6\doteq\omega_k\sim E_{\rm max}\).

The leading-order (LO) term of the expansion reproduces the purely local limit of the corrections, while the first next-to-local term provides the leading genuinely non-local contribution. Together with the local terms, these contributions account for the expansion through $\mathcal{O}(E_{\max}^{-3})$, as summarized in Tab.~\ref{tab:htet-bookkeeping}; accordingly, this level of approximation is referred to as next-to-leading order (NLO) in the $1/E_{\max}$ expansion. Both the LO and NLO approximations have already been derived in the literature~\cite{Cohen2022,Demiray:2025zqh}. In the present work, we focus on the NNLoc contributions.
At this order, the non-local factor appearing in the second line of Eq.~\eqref{eq:40legs} can be rewritten in terms of a universal symmetrized kernel for the four-leg sector. Specifically, once all four-leg topologies in Eq.~\eqref{diagram:all4legs_abcd} and their inequivalent symmetrizations are consistently included, the result becomes independent of the particular assignment of the external legs and can be expressed as

\begin{align}
    \label{eq:all_terms4legs}
    \mathcal{K}^4_{\mathrm{sym}}(E_f,E_i,\Omega_4,\omega_k; E_{\max})=&-\frac{2\Omega_4 +(E_f-E_i)^2+6\omega_k(E_f-E_i)+24\omega_k^2}
    {32\omega_k^5}\,\Theta(2\omega_k-E_{\rm max}) \notag\\
    &-\frac{-2\Omega_4 -(E_f-E_i)^2+6\omega_k(E_f+E_i)+3E_f(E_f-E_i)}
    {16\omega_k^4}\,\Theta'(2\omega_k-E_{\rm max})\notag\\
    &-\frac{2\Omega_4 +(E_f-E_i)^2+6E_fE_i}
    {16\omega_k^3}\,\Theta''(2\omega_k-E_{\rm max}),
\end{align}
where \(\Theta'(2\omega_k-E_{\rm max})\), \(\Theta''(2\omega_k-E_{\rm max})\) are, respectively, the first and second  derivatives of the Heaviside distribution, and \(\Omega_4\doteq \omega_1^{2}+\omega_2^{2}+\omega_3^{2}+\omega_4^{2}\).
We emphasize that the kernel \(\mathcal{K}^4_{\mathrm{sym}}\) exhibits a clear hierarchy in its dependence on the external energies.
At LO it is completely independent of \(E_i\) and \(E_f\).
At NLO it depends on the total energies of the initial and final states, but not on the individual external frequencies.
At NNLoc it acquires an irreducible dependence on the external frequencies, which enters only through the quadratic combination \(\Omega_4\).

To make the operator content explicit, consider the general expansion of
Eq.~\eqref{eq:all_terms4legs} at order \(n\). Once all four-leg topologies are accounted for, one
generically encounters monomials \(E_f^{m}E_i^{l}\) with \(m+l\le n\), which translate into
the operators
\begin{equation}
    \label{eq:operatorial_structure}
    E_f^{m}E_i^{l}
    \;\longrightarrow\;
     H_0^{m}\,\n{\phi^{4}}\,H_0^{l}\,.
\end{equation}
In addition, the \(\Omega_4\)-dependence forces the inclusion 
of further operator
structures. Considering again, as an explicit example, the topology in
Eq.~\eqref{eq:40legs}, one finds
\begin{align}
\label{eq:omega_factor}
&\sum_{1,2,3,4}\delta_{12,34}
(\omega_{1}^{2}+\omega_{2}^{2}+\omega_{3}^{2}+\omega_{4}^{2})
\bra{f}\,\phi^-_{4}\phi^-_{3}\phi^-_{2}\phi^-_{1}\,\ket{i}
\notag\\
&=
\frac{1}{2\pi R}
\sum_{k_1,k_2,k_3,k_4}
\delta_{k_1k_2,k_3k_4}
\frac{\omega_{k_1}^2+\omega_{k_2}^2+\omega_{k_3}^2+\omega_{k_4}^2}
{\sqrt{2\omega_{k_1}}\sqrt{2\omega_{k_2}}\sqrt{2\omega_{k_3}}\sqrt{2\omega_{k_4}}}
\bra{f}\, a^\dagger_{k_4} a^\dagger_{k_3} a^\dagger_{k_2} a^\dagger_{k_1}\,\ket{i}.
\end{align}
All these operators must therefore be incorporated into the effective Hamiltonian as
independent operator contributions, each multiplied by the corresponding numerical
coefficients determined by the matching procedure.

Factors of \(\omega_p\) multiplying the fields admit a natural interpretation in the interaction picture. Indeed, the mode operators obey
\begin{equation} 
\partial_t \phi_p^{\pm} = \mp i\,\omega_p\,\phi_p^{\pm} = i[H_0,\phi_p^{\pm}] \, , 
\end{equation} 
so that insertions of a single power of the frequency can be equivalently represented either as time derivatives acting on the fields or as commutators with the free Hamiltonian.
Since the dependence on the external frequencies enters the kernel \(\mathcal{K}^4_{\mathrm{sym}}\) only through the combination \(\Omega_4\), in which all frequencies appear quadratically, it is naturally associated with second-order time derivatives, or equivalently with double commutators with \(H_0\). Indeed, one finds
\begin{equation} 
\omega_p^2\,\phi_p^{\pm} = -\,\partial_t^2 \phi_p^{\pm} = [H_0,[H_0,\phi_p^{\pm}]] \, , 
\end{equation} 
independently of the sign of the frequency. This structure reflects the emergence of higher-derivative operator contributions at NNLoc.

Diagrams with two external legs can be treated in complete analogy. They are classified
into the following four topologies
\begin{align}
    \label{eq:all2legs_abcd}
     \adjustbox{scale=0.4, valign=c}{
\begin{tikzpicture}[line width=1.5pt]

\def\s{1.5}
\begin{feynman}
    \vertex (b);
    \vertex [left=\s cm of b] (a1) {$\scalemath{2.5}{a}$};
    \vertex (c)[right=\s*1.25 cm of b];
    \vertex [right=\s cm of c] (a2) {$\scalemath{2.5}{b}$};

    \adjustbox{scale=0.8, valign=c}{\includestandalone{Diagrams/*}} {
        (a1) -- (b),
        (b) -- [half left] (c) -- [half left] (b),
        (b) -- (c),
        (a2) -- (c),
    };

    \draw[fill=black] (b) circle (\s*1.5 pt);
    \draw[fill=black] (c) circle (\s*1.5 pt);
    \draw[draw=white] (-1.05*\s,-1.05*\s) rectangle (2.35*\s,1.05*\s);
\end{feynman}
\end{tikzpicture}
} \quad\quad
     \adjustbox{scale=0.4, valign=c}{
\begin{tikzpicture}[line width=1.5pt]

\def\s{1.5}
\begin{feynman}
    \vertex (b);
    \vertex (c)[right=\s*1.25 cm of b];

    \vertex [below=\s cm of c] (aux2);
    \vertex [right=\s cm of aux2] (a1) {$\scalemath{2.5}{b}$}; 
    
    \vertex [above=\s cm of b] (aux);
    \vertex [left=\s cm of aux] (a2) {$\scalemath{2.5}{a}$}; 
   
    \adjustbox{scale=0.8, valign=c}{\includestandalone{Diagrams/*}} {
        (a1) -- [in=-90, out=180] (b),
        (b) -- [half left] (c) -- [half left] (b),
        (b) -- (c),
        (a2) -- [in=90, out=0] (c),
    };

    \draw[fill=black] (b) circle (\s*1.5 pt);
    \draw[fill=black] (c) circle (\s*1.5 pt);
    \draw[draw=white] (-1.05*\s,-1.05*\s) rectangle (2.35*\s,1.05*\s);
\end{feynman}
\end{tikzpicture}
} \quad\quad
     \adjustbox{scale=0.4, valign=c}{
\begin{tikzpicture}[line width=1.5pt]

\def\s{1.5}
\begin{feynman}
    \vertex (b);
    \vertex [left=\s cm of b] (a1) {$\scalemath{2.5}{b}$};
    \vertex (c)[right=\s*1.25 cm of b];

    \vertex [above=\s cm of b] (aux);
    \vertex [left=\s cm of aux] (a2) {$\scalemath{2.5}{a}$}; 

    \vertex [below=\s cm of b] (aux2);
    \vertex [left=\s cm of aux2] (a3) {$\scalemath{2.5}{\phantom{b}}$}; 

    \adjustbox{scale=0.8, valign=c}{\includestandalone{Diagrams/*}} {
        (a1) -- (b),
        (b) -- [half left] (c) -- [half left] (b),
        (b) -- (c),
        (a2) -- [in=90, out=0] (c),
    };

    \draw[fill=black] (b) circle (\s*1.5 pt);
    \draw[fill=black] (c) circle (\s*1.5 pt);
    \draw[draw=white] (-1.05*\s,-1.05*\s) rectangle (1.35*\s,1.05*\s);
\end{feynman}
\end{tikzpicture}
} \quad\quad
     \adjustbox{scale=0.4, valign=c}{
\begin{tikzpicture}[line width=1.5pt]

\def\s{1.5}
\begin{feynman}
    \vertex (b);
    \vertex (c)[right=\s*1.25 cm of b];
    \vertex [right=\s cm of c] (a1) {$\scalemath{2.5}{b}$};
    \vertex [above=\s cm of c] (aux);
    \vertex [right=\s cm of aux] (a2) {$\scalemath{2.5}{a}$}; 

    \vertex [below=\s cm of c] (aux2);
    \vertex [right=\s cm of aux2] (a3) {$\scalemath{2.5}{\phantom{b}}$}; 

    \adjustbox{scale=0.8, valign=c}{\includestandalone{Diagrams/*}} {
        (a1) -- (b),
        (b) -- [half left] (c) -- [half left] (b),
        (b) -- (c),
        (a2) -- [in=90, out=180] (b),
    };

    \draw[fill=black] (b) circle (\s*1.5 pt);
    \draw[fill=black] (c) circle (\s*1.5 pt);
     \draw[draw=white] (-0.1*\s,-1.05*\s) rectangle (2.35*\s,1.05*\s);
\end{feynman}
\end{tikzpicture}
}
\end{align}
The resulting expressions mirror those of the four-leg case in
Eq.~\eqref{eq:all_terms4legs}. After summing over all topologies in
Eq.~\eqref{eq:all2legs_abcd} and performing the corresponding symmetrizations,
the NNLoc non-local corrections can be organized in terms of a universal
symmetrized kernel $\mathcal{K}^2_{\mathrm{sym}}$, whose structure is independent
of the specific external-leg assignment.
The associated operatorial content follows from the
mapping of energy factors according to
\begin{equation}
    E_f^{m}\,E_i^{l}
    \;\longrightarrow\;
    H_0^{m}\,\n{\phi^{2}}\,H_0^{l}\,.
\end{equation}
At NNLoc, the two-leg sector  develops 
the same non-local feature as in the four-leg case, namely the appearance of additional contributions 
proportional to \(\Omega_2 \doteq \omega_1^{2} + \omega_2^{2}\).
Again, this encodes an irreducible dependence on the
individual external frequencies, which enter the kernel
$\mathcal{K}^2_{\mathrm{sym}}$ exclusively through a quadratic combination, in
direct analogy with the role played by $\Omega_4$.

The case with no external legs is simpler, as only the following topology contributes
\begin{equation}
    \label{eq:all0legs_abcd}
    \adjustbox{scale=0.4, valign=c}{
\begin{tikzpicture}[line width=1.5pt]

\def\s{1.5}
\begin{feynman}
    \vertex (b);
    \vertex (c)[right=\s*1.25 cm of b];

    \adjustbox{scale=0.8, valign=c}{\includestandalone{Diagrams/*}} {
        (b) -- [half left] (c) -- [half left] (b),
        (b) -- [half left, looseness=0.5] (c)
             -- [half left, looseness=0.5] (b),
    };

    \draw[fill=black] (b) circle (\s*1.5 pt);
    \draw[fill=black] (c) circle (\s*1.5 pt);
    \draw[draw=white] (-0.1*\s,-0.6*\s) rectangle (1.35*\s,0.6*\s);
\end{feynman}
\end{tikzpicture}
}
\end{equation}
The corresponding operatorial contributions involve only powers of the free Hamiltonian, according to
\begin{equation}
\label{eq:mapzerolegs}
     E_{i}^{m+l}=E_{f}^{m+l}
    \;\longrightarrow\;
    \bra{f}\, H_0^{m+l}\,\ket{i},
\end{equation}

In order to determine the numerical coefficients multiplying the various operatorial structures discussed above, one must examine the distributional contributions arising in the expansion of the matching expressions. At order \(n\) in this expansion, terms proportional to the \(n\)-th derivative of the Heaviside step function typically appear. Considering the four–external–leg case in Eq.~\eqref{eq:all_terms4legs}, these contributions schematically take the form
\begin{equation}
 \label{eq:problem_Heaviside}
 \sum_k \frac{1}{\omega_k^\alpha}\,\frac{d^n}{d(2\omega_k)^n}\,\Theta(2\omega_k-E_{\rm max})\,.
\end{equation}
In finite volume the spectrum is discrete, therefore distributions such as \(\delta(2\omega_k-E_{\rm max})\) (and higher derivatives of \(\Theta\) as well) do not correspond to a unique numerical prescription: if one tries, e.g., to assign the prescription
\begin{equation}
\delta(2\omega_k-E_{\rm max}) = \begin{cases}
    1 & \ {\rm if }\  2\omega_k=E_{\rm max}\\
    0 & \ {\rm  otherwise}
\end{cases}\;,
\end{equation}
the resulting contribution vanishes for generic values of \(E_{\rm max}\), \(R\), and \(m\), since the condition \(2\omega_k = E_{\rm max}\) requires
\[
k = R\;\sqrt{\frac{E_{\rm max}^2}{4}-m^2}
\]
to be an integer, which is not guaranteed for generic choices of the parameters.

This issue is not intrinsic to the matching procedure, but rather an artifact of performing the
\(1/E_{\rm max}\) expansion after the spectrum discretization. Formulating the matching in infinite volume provides a natural way to avoid it, as the spectrum becomes continuous and the distributional structure associated
with derivatives of the Heaviside step function is well defined in the standard sense. Once the coefficients are determined in the continuum, the discretization can be reintroduced through spatial compactification, which is required in practice to obtain a separable Hilbert space, i.e.,  a space admitting a discrete basis.

In infinite volume, the sum in Eq.~\eqref{eq:problem_Heaviside} is replaced by a momentum integral,
leading to
\begin{equation}
 \label{eq:Integral_F}
 F_n^\alpha(E_{\rm max})
 \doteq
 \int_{-\infty}^{\infty}\mathrm{d}k\;
 \frac{1}{\omega(k)^\alpha}\,
 \frac{d^n}{d(2\omega(k))^n}\,
 \Theta\!\bigl(2\omega(k)-E_{\rm max}\bigr).
\end{equation}
The discussion above can be straightforwardly extended to the cases with two and zero external legs. In these sectors, the determination of the matching coefficients similarly reduces to the evaluation of continuum integrals involving derivatives of the Heaviside step function, with the only difference being the structure of the energy denominators associated with the corresponding topologies.
 In the two-leg case, the relevant integrals take the form
\begin{equation}
\label{eq:integral_I}
 I_n^\alpha (E_{\rm max})
 \doteq
 \int_{-\infty}^{\infty} \mathrm{d}k
 \int_{-\infty}^{\infty} \mathrm{d}q\;
 \frac{1}{P_3(k,q)\,W_3^\alpha(k,q)}
 \frac{d^n}{dW_3(k,q)^n}
 \Theta\!\bigl(W_3(k,q)-E_{\rm max}\bigr),
\end{equation}
where we have defined
\begin{equation}
    \label{eq:def_P2_W2}
    P_3(k,q) \doteq \omega(k)\,\omega(q)\,\omega(k+q),
    \quad\quad
    W_3(k,q) \doteq \omega(k)+\omega(q)+\omega(k+q).
\end{equation}
Similarly, diagrams with no external legs give rise to
\begin{equation}
    \label{eq:integral_B}
    B_n^\alpha (E_{\rm max})
    \doteq
    \int_{-\infty}^{\infty} \mathrm{d}k
    \int_{-\infty}^{\infty} \mathrm{d}p
    \int_{-\infty}^{\infty} \mathrm{d}q\;
    \frac{1}{P_4(k,p,q)\,W_4^\alpha(k,p,q)}
    \frac{d^n}{dW_4(k,p,q)^n}
    \Theta\!\bigl(W_4(k,p,q)-E_{\rm max}\bigr),
\end{equation}
with
\begin{equation}
    \label{eq:def_P0_W0}
    P_4(k,p,q)
    \doteq
    \omega(k)\,\omega(p)\,\omega(q)\,\omega(k+p+q),
    \quad\quad
    W_4(k,p,q)
    \doteq
    \omega(k)+\omega(p)+\omega(q)+\omega(k+p+q).
\end{equation}
The explicit evaluation of the integrals \(F_n^\alpha\), \(I_n^\alpha\), and \(B_n^\alpha\) is technically non-trivial, due to the presence of derivatives of the Heaviside step function and the non-linear dependence of the integrands on the integration variables.
The calculation is therefore deferred to App.~\ref{app:integral_computation}, where we describe the general strategy and provide the details of the derivation.

In Ref.~\cite{Demiray:2025zqh}, the first non-local coefficients,
\(\xi, \alpha_1^{(1)}, \alpha_2^{(2)}, \beta_1^{(1)}, \beta_2^{(2)}\),
are derived; the corresponding expressions are reported  in App.~\ref{app:NLO} for completeness.

Their determination relies on different practical prescriptions. 
Some coefficients are computed directly in infinite volume, for example \(\beta^{(1)}_1\), while others are evaluated as discrete sums over finite-volume modes, such as \(\beta^{(2)}_2\).

These definitions coincide in the infinite-volume limit. 
However, at the finite volumes relevant for numerical implementations, they can yield quantitatively different results, as discussed in App.~\ref{app:finite_volume}. 
To have a single and consistent definition for all coefficients, we adopt an infinite-volume matching.

\begin{figure}[t]
    \centering
      \includegraphics[width=0.7\textwidth]{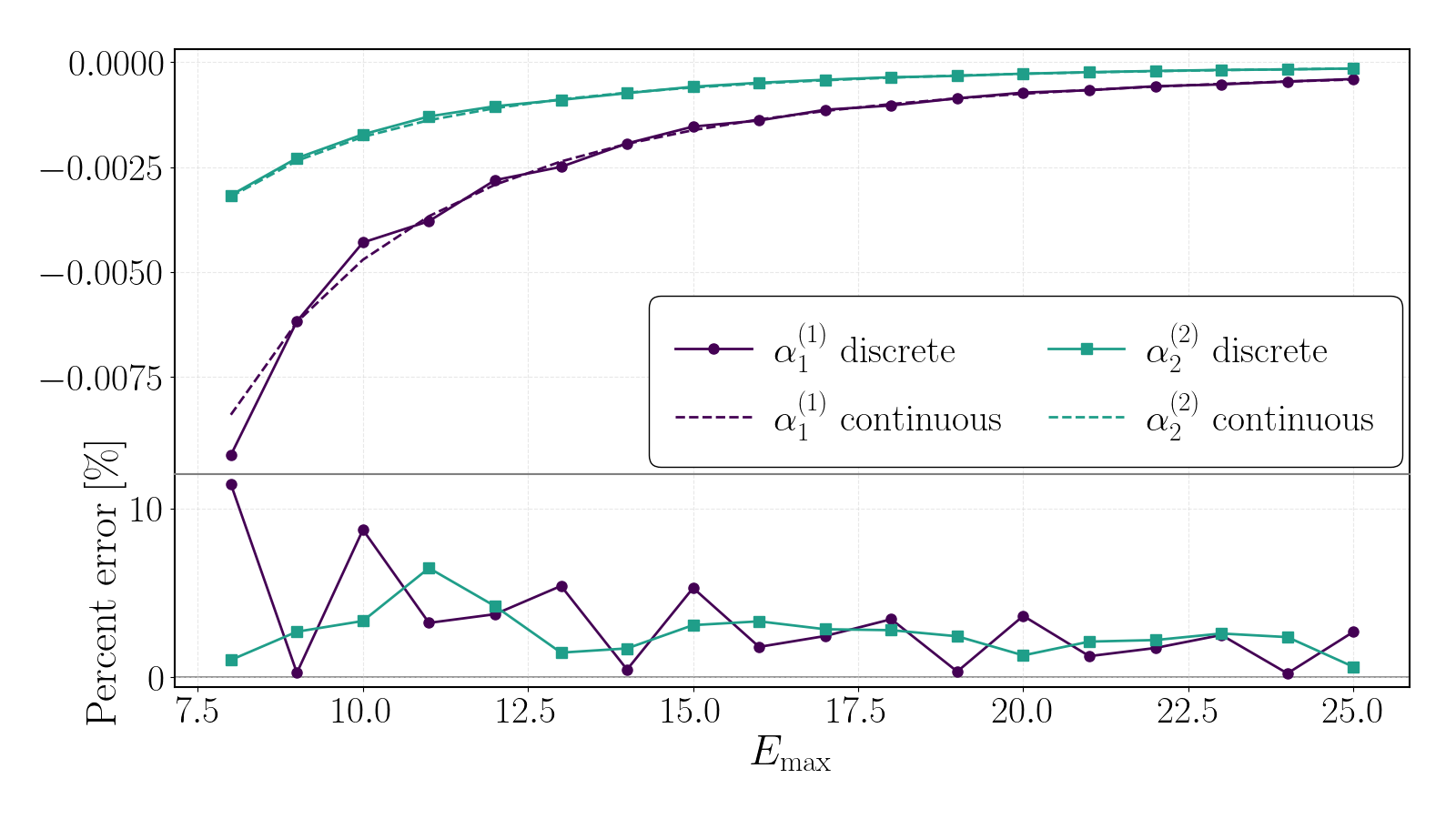}
        \includegraphics[width=0.7\textwidth]{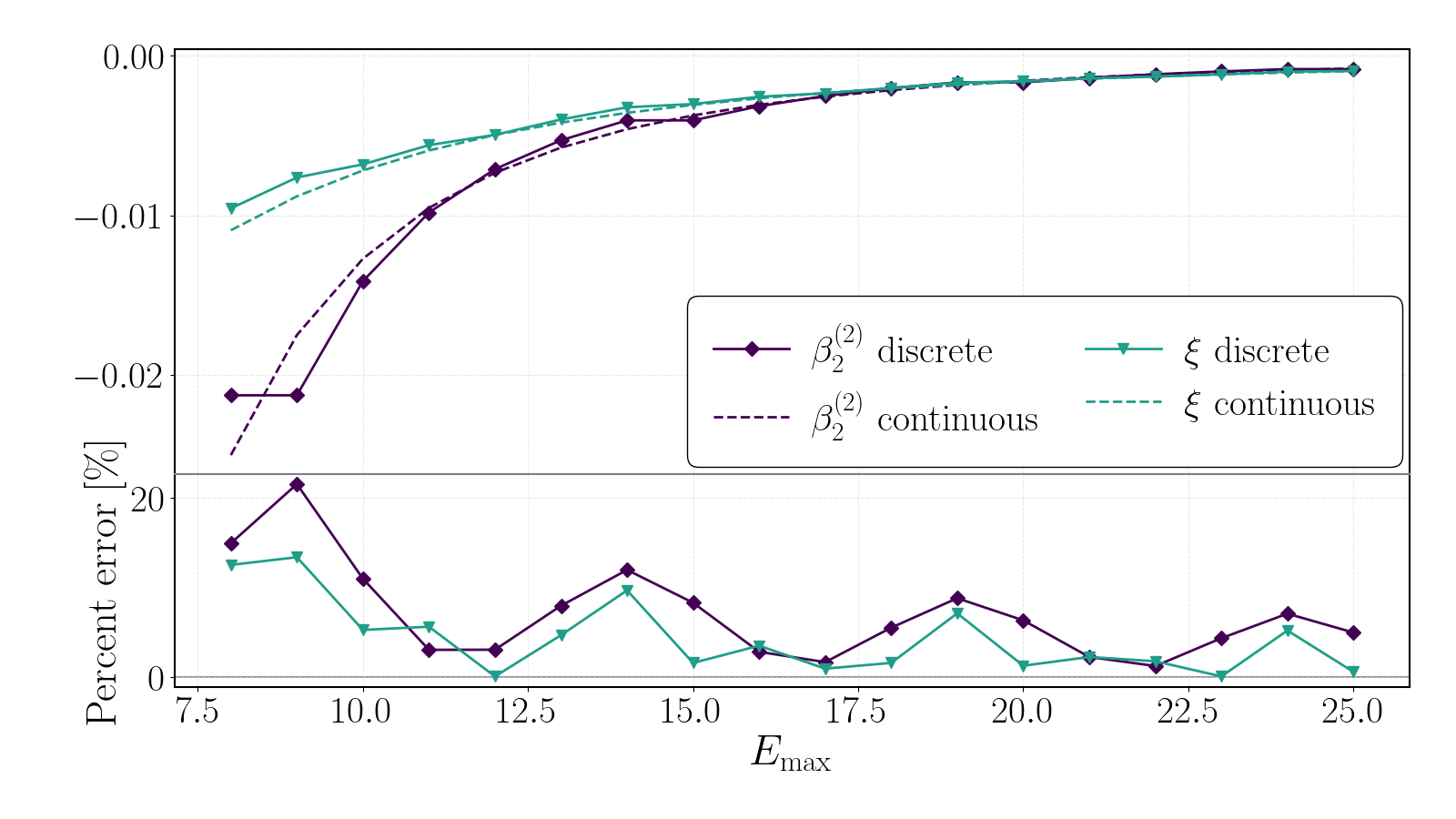}
\caption{
First non-local corrections \(\xi\), \(\alpha_1^{(2)}\), \(\alpha_2^{(2)}\), and \(\beta_2^{(2)}\), computed at finite (solid line) and infinite (dotted line) volume as a function of the energy cut-off \(E_{\max}\), for \(2\pi R = 10\), \(\lambda/4\pi = 1\) and \(m=1\), in the range \(E_{\max}\in[8,25]\).
The upper panel shows the corrections \(\alpha_1^{(2)}\) and \(\alpha_2^{(2)}\), while the lower panel shows \(\xi\) and \(\beta_2^{(2)}\). In both panels, the lower portion of each plot shows the corresponding percent error with respect to the continuum value.
For the discrete corrections, the sums over momenta are evaluated up to a finite maximum momentum \(K_{\mathrm{UV}}=1000\).
}
\label{fig:Correctiondis_vs_cont}
\end{figure}
Figure~\ref{fig:Correctiondis_vs_cont} shows the ratio between the leading non-local corrections computed at finite and infinite volume as a function of the energy cut-off \(E_{\max}\). At low \(E_{\max}\), the discrete evaluation fluctuates around the continuum result, indicating the presence of finite-volume effects.
For larger values of \(E_{\max}\), the discrete results instead converge to the continuum corrections, which exhibit a smooth and stable behavior as \(E_{\max}\) is varied.

Having determined the matching coefficients in infinite volume, we can now address the problem of implementing the resulting operatorial structures in a separable Hilbert space. This requires compactifying the spatial direction, which discretizes the spectrum and makes it possible to evaluate the operators on a finite truncated basis. The compact spatial geometry is therefore introduced only at this stage, as the matching calculation itself does not require a discrete spectrum.

The effective Hamiltonian thus contains contributions that combine continuum matching
coefficients with operators defined in finite volume.
Table~\ref{tab:operators_corrections} summarizes the results, presenting each operator
together with its corresponding correction, where we have introduced the operators
\(H_2\) and \(H_4\) as
\begin{equation}
    \label{eq:def_H_n}
    H_2 \doteq \int R\,\mathrm{d}\theta\;\n{\phi^2}\;, \quad H_4 \doteq \int R\,\mathrm{d}\theta\;\n{\phi^4}\;.
\end{equation}
The operators and their associated corrections are organized according to their scaling
in inverse powers of the truncation scale \(E_{\rm max}\), from which we can confirm that
the second non-local corrections scale as \(\mathcal{O}(E_{\rm max}^{-4})\).

\begin{table}[ht]

\centering
\setlength{\tabcolsep}{10pt}
\setlength{\arrayrulewidth}{0.6pt}

\begin{tabular}{c c|c c}


\hhline{--|--}
\multicolumn{2}{c|}{\textbf{Lower orders}} &
\multicolumn{2}{c}{\textbf{Higher order}} \\

Operator & Coefficient & Operator & Coefficient \\
\hhline{--|--}


\multicolumn{2}{c|}{$\mathcal{O}(E_{\rm max}^{-2})$}\Ttabular\Btabular &
\multicolumn{2}{c}{$\mathcal{O}(E_{\rm max}^{-4})$} \\

\hhline{--|--}

$I$ \Ttabular\Btabular
& $-\dfrac{\lambda^2 R}{1536\pi^2} B_0^1$
& $H_0^2$
& $-\dfrac{\lambda^2 R}{768(2\pi)^2} B_2^1$ \\

$H_2$\Ttabular\Btabular
& $-\dfrac{\lambda^2}{96\pi^2} I_0^1$
& $H_0^2 H_2$
& $-\dfrac{\lambda^2}{192\pi^2} I_1^2$ \\

$H_4$\Ttabular\Btabular
& $-\dfrac{3\lambda^2}{16\pi} F_0^3$
& $H_0^2 H_4$
& $-\dfrac{\lambda^2}{128\pi}(F_0^5 + 4F_1^4 + 2F_2^3)$ \\

\hhline{--|~~}
\multicolumn{2}{c|}{$\mathcal{O}(E_{\rm max}^{-3})$} \Ttabular\Btabular&
$H_4 H_0^2$
& $-\dfrac{\lambda^2}{128\pi}(F_0^5 - 2F_1^4 + 2F_2^3)$ \\
\hhline{--|~~}

$H_0$\Ttabular\Btabular
& $-\dfrac{\lambda^2 R}{384(2\pi)^2} B_1^1$
& $H_0 H_2 H_0$
& $-\dfrac{\lambda^2}{192\pi^2}(-I_1^2 + I_2^1)$ \\

$H_0 H_2$\Ttabular\Btabular
& $-\dfrac{\lambda^2}{192\pi^2}(I_0^2 + I_1^1)$
& $H_0 H_4 H_0$
& $-\dfrac{\lambda^2}{64\pi}(-F_0^5 - F_1^4 + 4F_2^3)$ \\

$H_0 H_4$\Ttabular\Btabular
& $-\dfrac{3\lambda^2}{64\pi}(F_0^4 + 2F_1^3)$
& $\sim \Omega_2 H_2$
& $-\dfrac{\lambda^2}{384\pi^2}(2I_0^3 - 2I_1^2 + I_2^1)$ \\

$H_2 H_0$\Ttabular\Btabular
& $-\dfrac{\lambda^2}{192\pi^2}(-I_0^2 + I_1^1)$
& $\sim \Omega_4 H_4$
& $-\dfrac{\lambda^2}{64\pi}(F_0^5 - 2F_1^4 + 2F_2^3)$ \\

$H_4 H_0$\Ttabular\Btabular
& $-\dfrac{3\lambda^2}{64\pi}(-F_0^4 + 2F_1^3)$
& & \\

\hhline{--|--}
\end{tabular}

\caption{Operators organized according to their scaling in inverse powers of the truncation scale
\(E_{\rm max}\). Lower-order contributions are shown in the left column, while the
\(\mathcal{O}(E_{\rm max}^{-4})\) terms associated with the second non-local corrections
are collected in the right column. When entering the effective Hamiltonian, the
contributions involving \(H_4\) and \(H_2\) must be divided by the appropriate symmetry
factors, namely \(4!\) and \(2\), respectively.}
\label{tab:operators_corrections}
\end{table}

In the zero-external-leg sector, the NNLoc operatorial structure involves only powers of the free Hamiltonian \(H_0\), see Eq.~\eqref{eq:mapzerolegs}. Moreover, the coefficients arising from derivatives of the Heaviside distribution can always be rewritten as derivatives with respect to the cut-off (see Eq.~\eqref{eq:derivative_Theta} in App.~\ref{app:integral_computation}).
At order \(n\), the corresponding numerical prefactor scales as \(1/n!\). As a result, the full tower of non-local contributions in the vacuum sector can be resummed into a compact operator, i.e.,
\begin{align}
\sum_{n=0}^{\infty} \frac{1}{n!}\, H_0^{\,n}\,(-\partial_{E_{\max}})^{n} B^{1}_{0}(E_{\max})
&= e^{-H_0\,\partial_{E_{\max}}}\, B^{1}_{0}(E_{\max})
= B^{1}_{0}(E_{\max}-H_0)\,,
\label{eq:vacuum_exponentiation}
\end{align}
where the last equality is defined through the Taylor expansion of 
$B^{1}_{0}(E_{\max}-H_0)$. 
Equivalently, when acting on an eigenstate of \(H_0\)  with energy \(\mathcal{E}\), it  yields \(B^{1}_{0}(E_{\max}-\mathcal{E})\).
This shows that, in the vacuum sector, the non-local \(1/E_{\max}\) expansion resums exactly to an operator-valued function of \(H_0\).

\section{Numerical Results}
\label{sec_Numerical Results}
In this section we implement the resummed matching corrections to the quartic coupling and to the quadratic operator,
given in Eqs.~\eqref{eq:delta_lambda_resum} and \eqref{eq:delta_m_resum} and derived in Sec.~\ref{sec_resummation_of_local_expansion},
together with the NNLoc non-local operator insertions summarized in Tab.~\ref{tab:operators_corrections} and discussed in Sec.~\ref{sec_NNLO_corrections}.

For each value of the truncation scale \(E_{\max}\) we construct the truncated Hilbert space \(\mathcal{H}_{\mathrm{tru}}\) and diagonalize the
corresponding Hamiltonian matrix. 
The database containing all eigenvalues extracted from the effective Hamiltonians, for the various prescriptions employed in the analysis of this section, can be found in Ref.~\cite{Github2026HTET}.

As a representative observable we consider the first spectral gap,
\begin{equation}
\Delta E_1(E_{\max}) \doteq E_1(E_{\max}) - E_0(E_{\max})\,,
\end{equation}
where \(E_0\) and \(E_1\) denote the lowest two eigenvalues of the effective Hamiltonian.

\begin{figure}[t]
  \centering
  \includegraphics[width=0.75\linewidth]{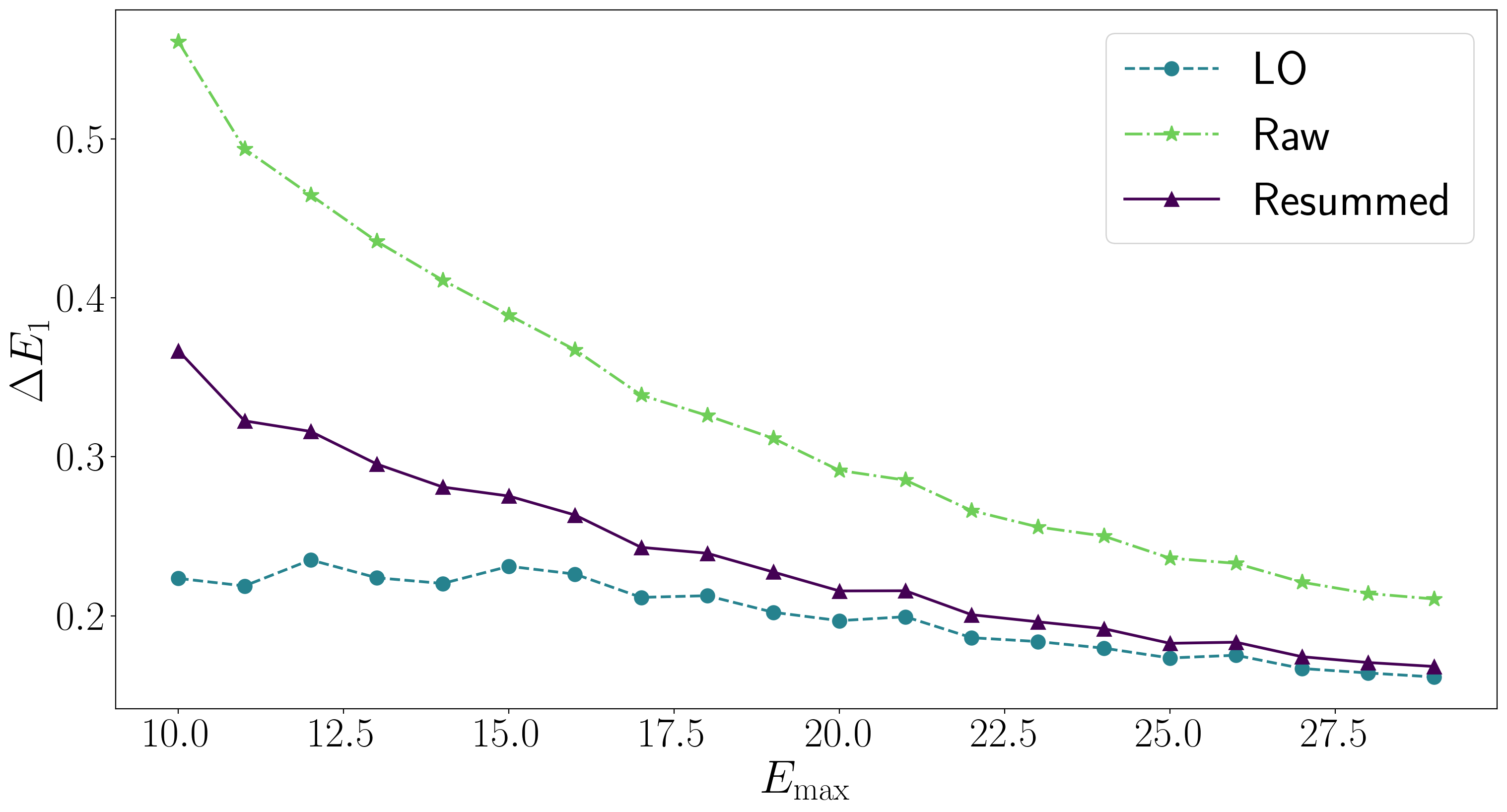}
\caption{First energy gap \(\Delta E_1\) as a function of the truncation scale \(E_{\max}\), at fixed parameters
\(m=1\), \(\lambda/(4\pi)=1\), and \(2\pi R=10\).
The curve  (dot-dashed lines with star markers) labeled \enquote{Raw} corresponds to the bare truncation \(H_{\mathrm{raw}}=PHP\).
The curve (dashed lines with circular markers) labeled \enquote{LO} includes the leading local counterterms.
The curve (solid lines with triangular markers) labeled \enquote{Resummed} includes the resummed local corrections implemented via Eqs.~\eqref{eq:delta_lambda_resum} and \eqref{eq:delta_m_resum}.}
\label{fig:gap_resummation}
\end{figure}

Figure~\ref{fig:gap_resummation} compares \(\Delta E_1\) obtained from three prescriptions:
(i) the raw truncation \(H_{\mathrm{raw}} \doteq P H P\);
(ii) the LO improvement, where only the lowest-order local counterterms are included;
(iii) the resummed local improvement, where the local counterterms are implemented using the all-order expressions
for \(\delta\tilde{\lambda}(E_{\max})\) and \(\delta\tilde{m}^2(E_{\max})\) from Sec.~\ref{sec_resummation_of_local_expansion}.

In the range of \(E_{\max}\) explored here, the \enquote{resummed} prescription does not exhibit a systematic improvement over the LO result. The underlying reason is that the resummation does not provide a uniformly improved approximation order by order: it resums to all orders only a fixed diagrammatic topology and only within the local approximation. In this regime, non-local contributions can be numerically comparable to the higher-order local terms that are being resummed, and additional diagrammatic topologies may yield significant contributions at the same order in the \(1/E_{\max}\) expansion. 

Resumming only the local contributions can lead to over-correction, which may slow the  convergence with 
\(E_{\max}\).  In contrast, the LO-improved Hamiltonian includes all local corrections at 
\(O(V^2)\) through a controlled perturbative matching, ensuring more reliable convergence.

This observation  motivates the inclusion of the non-local operator insertions discussed in Sec.~\ref{sec_NNLO_corrections} and summarized in Tab.~\ref{tab:operators_corrections}, which account for  the missing subleading contributions in the \(1/E_{\max}\) expansion.

\begin{figure}[t]
    \centering
      \includegraphics[width=0.7\textwidth]{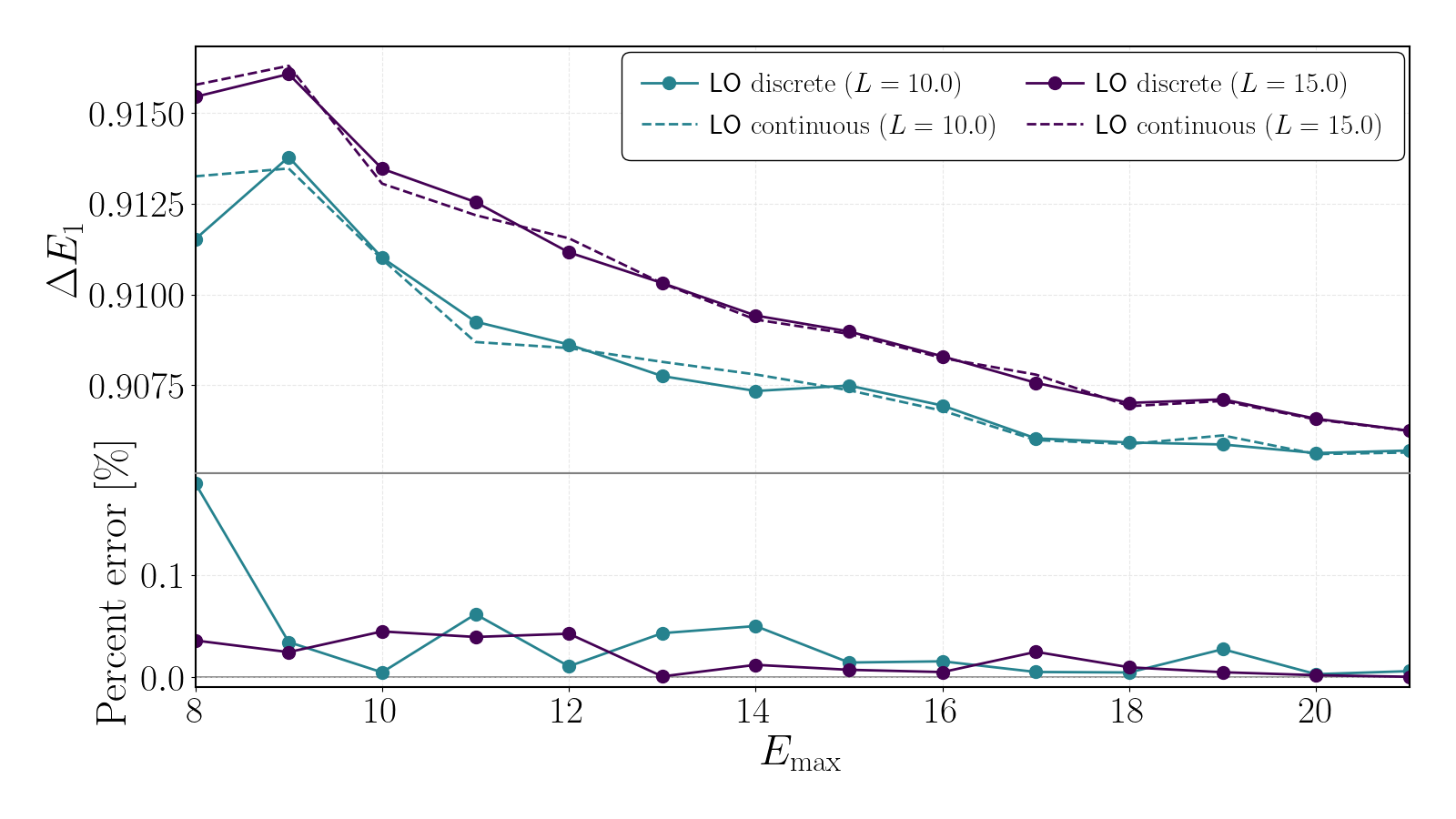}
        \includegraphics[width=0.7\textwidth]{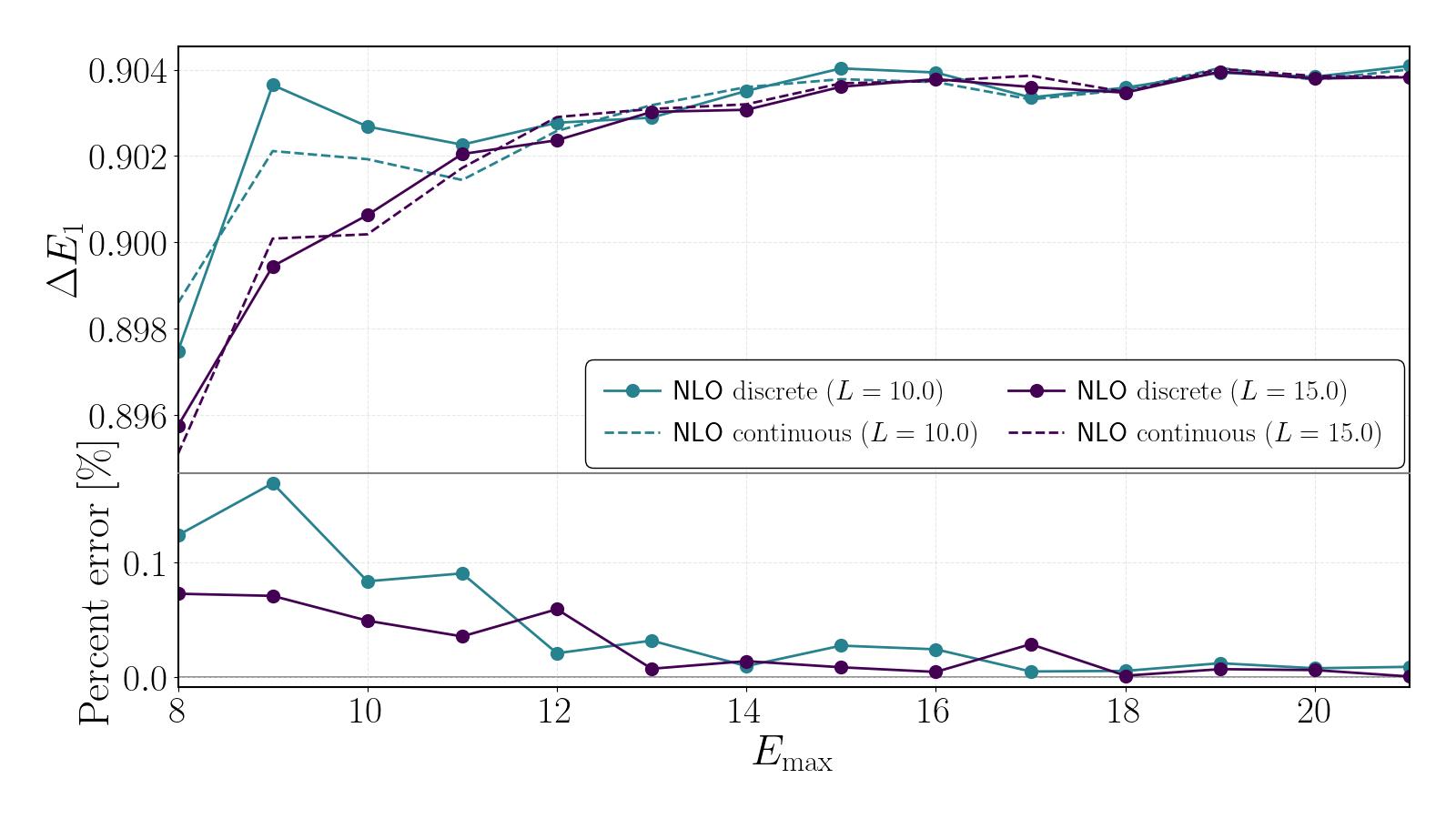}
\caption{First energy gap \(\Delta E_1\) as a function of the cut-off \(E_{\max}\) for two circumferences, \(L=10\) and \(L=15\), at fixed parameters \(m=1\) and \(\lambda/(4\pi)=1\). The upper panel shows the LO implementation (including corrections up to \(\mathcal{O}(E_{\max}^{-2})\)), while the lower panel shows the NLO implementation (including corrections up to \(\mathcal{O}(E_{\max}^{-3})\)). In each panel, the lower sub-plot reports the percent deviation of the discrete result from the corresponding continuum evaluation. We observe that, in the infinite-volume case, the dependence of the spectrum on the value of the circumference originates from the operatorial sector rather than from the coefficients.}
\label{fig:disvscont}
\end{figure}

Before addressing the NNLoc corrections, Fig.~\ref{fig:disvscont} compares the first energy gap \(\Delta E_1\) obtained from discrete (finite-volume) and continuous (infinite-volume) evaluations for two values of the circumference, \(L=10\) and \(L=15\), at LO and NLO.

Comparing the two orders shows that the discrepancy between the discrete and continuous treatments decreases as the cut-off \(E_{\max}\) is increased. Moreover, apart from isolated spikes, increasing the volume from \(L=10\) to \(L=15\) systematically suppresses residual finite-volume effects and results in faster and more uniform convergence to the continuum limit as \(E_{\max}\) increases.   

\begin{figure}[t]
  \centering

  \begin{minipage}[t]{0.5\linewidth}
    \centering
    \includegraphics[width=0.8\linewidth]{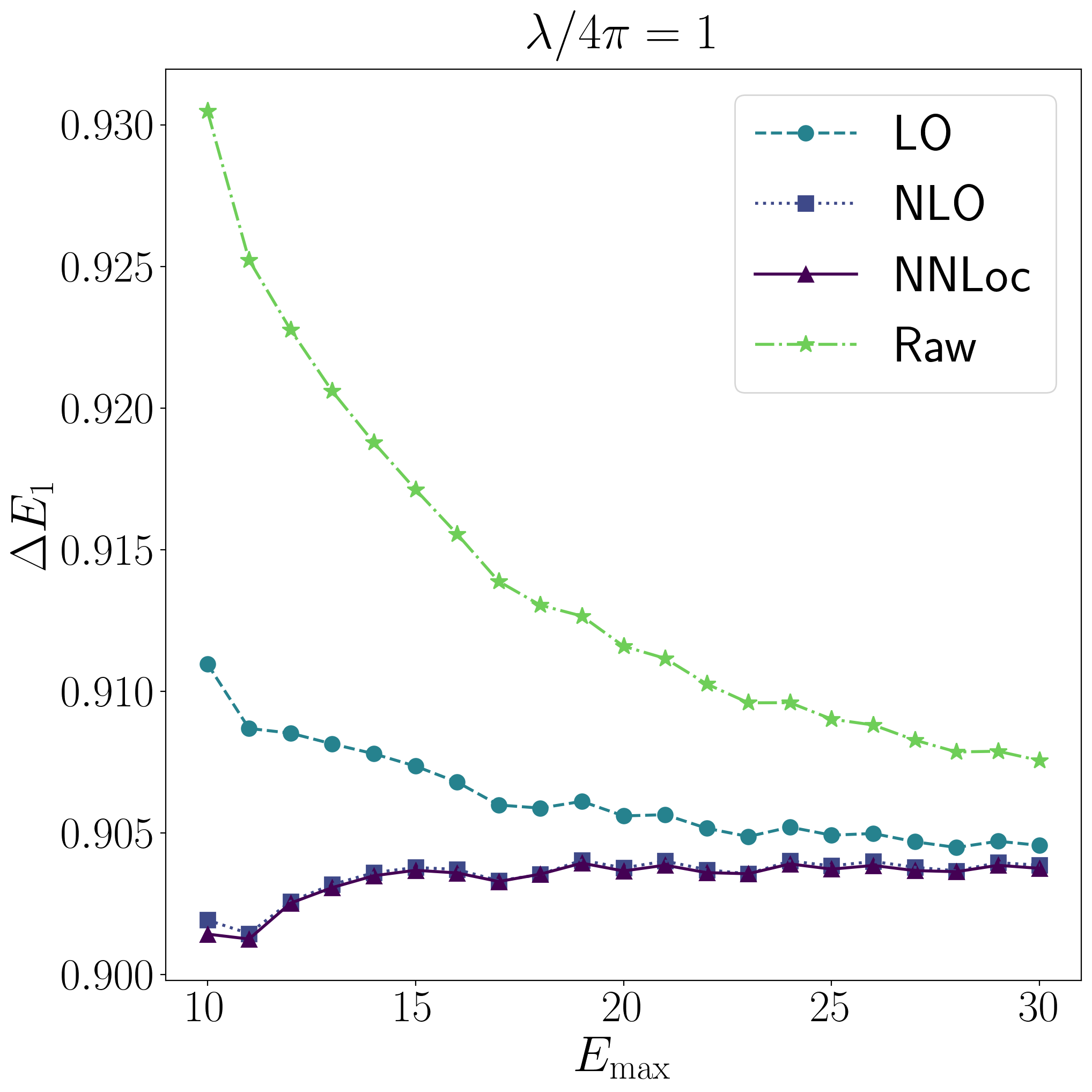}
  \end{minipage}\hfill
  \begin{minipage}[t]{0.5\linewidth}
    \centering
    \includegraphics[width=0.8\linewidth]{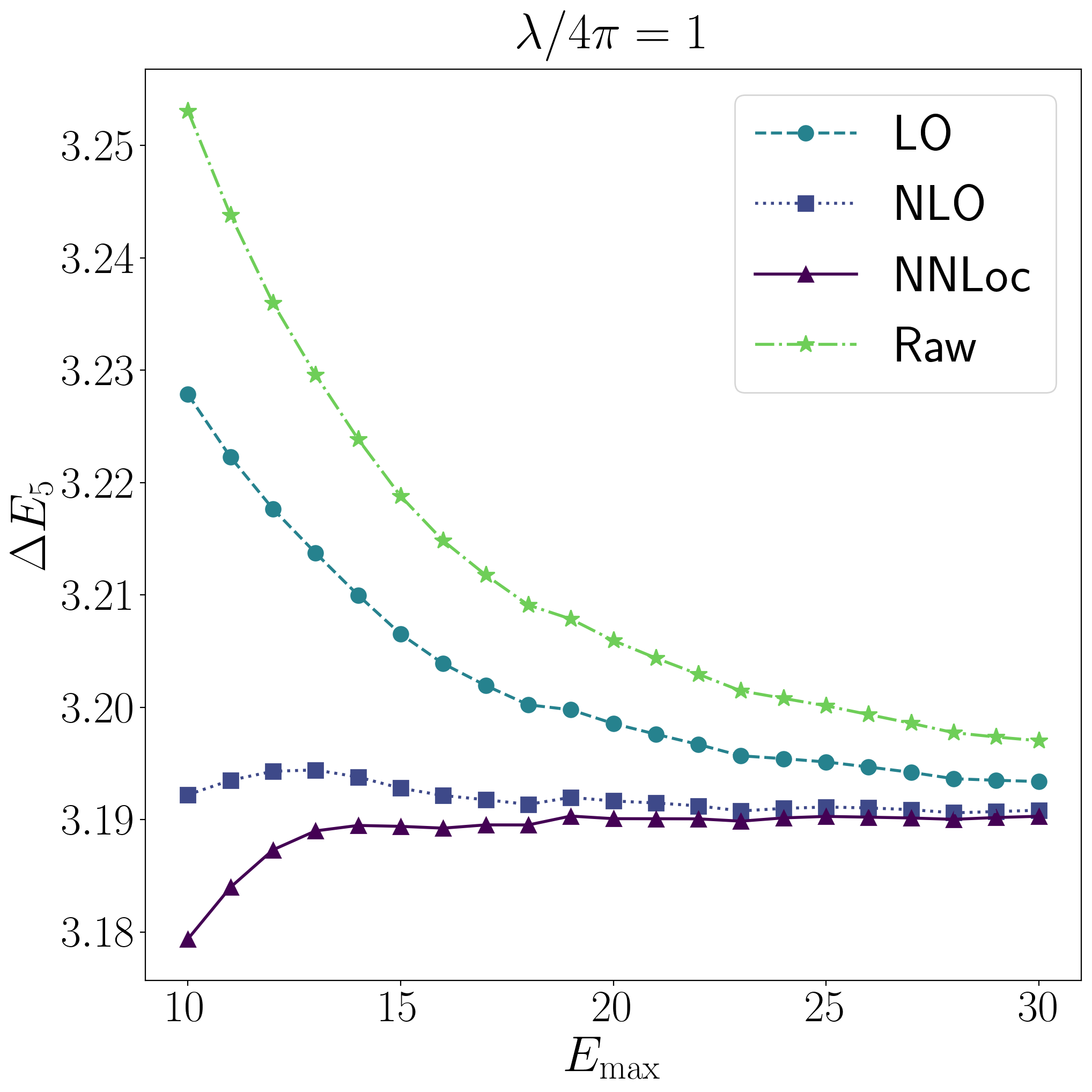}
  \end{minipage}

  \vspace{0.8em}

  \begin{minipage}[t]{0.5\linewidth}
    \centering
    \includegraphics[width=0.8\linewidth]{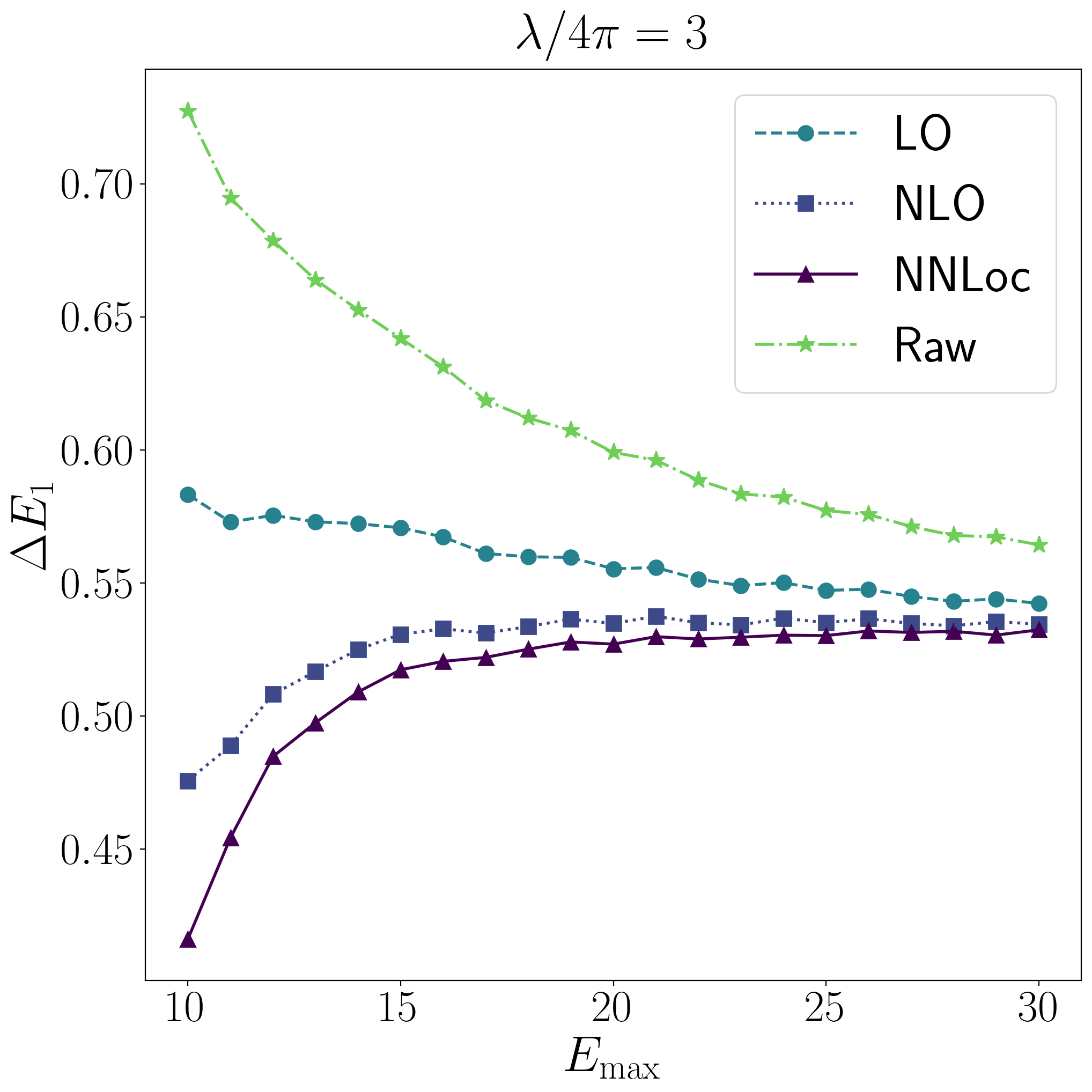}
  \end{minipage}\hfill
  \begin{minipage}[t]{0.5\linewidth}
    \centering
    \includegraphics[width=0.8\linewidth]{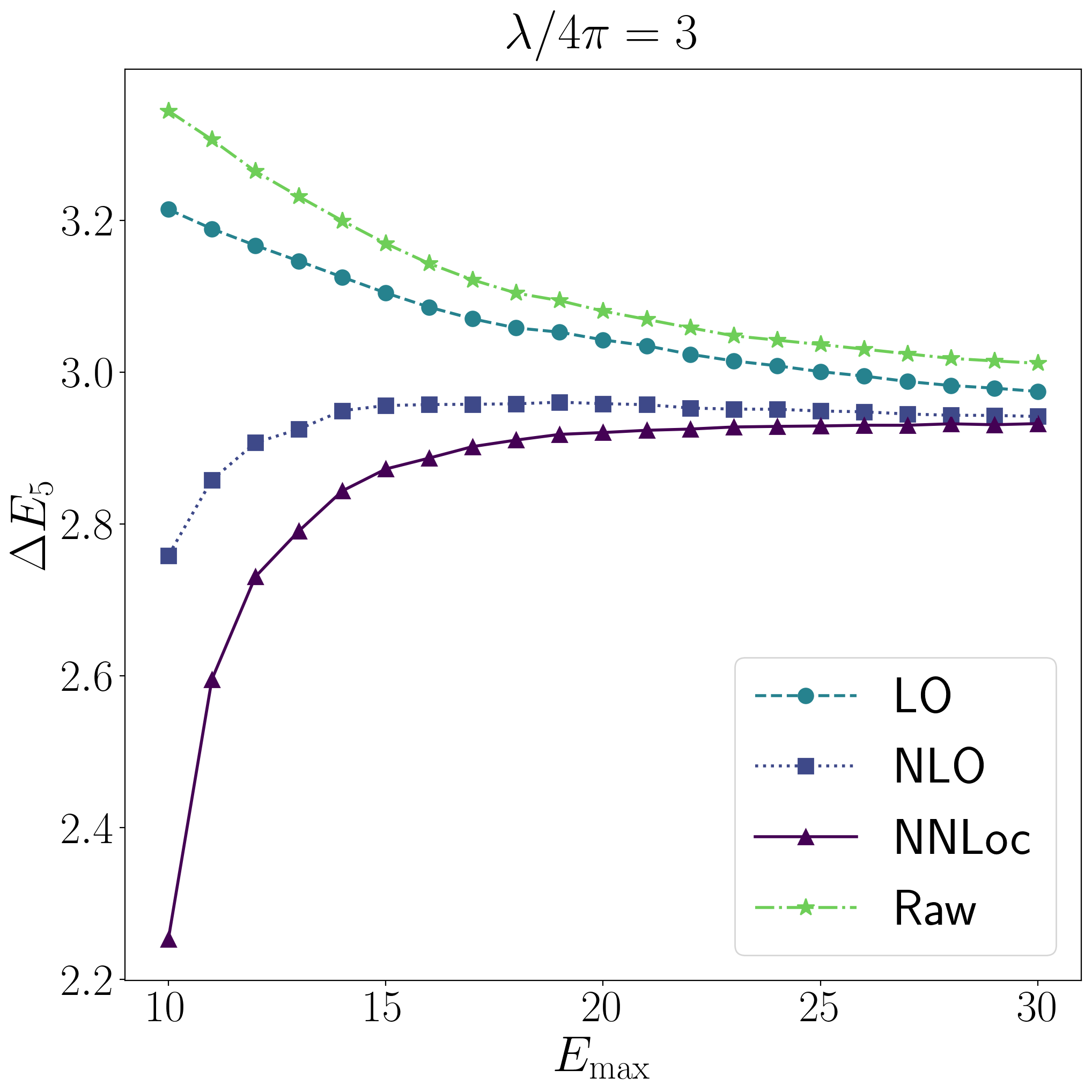}
  \end{minipage}

\caption{Dependence of the energy gaps on the cut-off \(E_{\max}\), at fixed parameters \(m=1\) and \(L=10\), comparing different HT and HTET prescriptions:
Raw (stars with dash-dotted lines), LO (circles with dashed lines), NLO (squares with dotted lines), and NNLoc (triangles with solid lines).
Clockwise from the top left: \(\Delta E_1\) with \(\lambda/(4\pi)=1\), \(\Delta E_5\) with \(\lambda/(4\pi)=1\),
\(\Delta E_5\) with \(\lambda/(4\pi)=3\), and \(\Delta E_1\) with \(\lambda/(4\pi)=3\).  }

  \label{fig:gaps_Emax_4panel}
\end{figure}

We conclude by presenting the numerical results for the NNLoc non-local corrections summarized in Tab.~\ref{tab:operators_corrections}. 
Figure~\ref{fig:gaps_Emax_4panel} shows the first energy gap \(\Delta E_1\) and  the higher gap \(\Delta E_5 \doteq E_5 - E_0\) as functions of $E_{\max}$, with \(\Delta E_5 \)
varying more smoothly with the cut-off.
Both observables are displayed for two values of the coupling, \(\lambda/(4\pi)=1\) and \(\lambda/(4\pi)=3\). For each configuration, all truncation prescriptions are implemented, namely Raw, LO, NLO, and NNLoc.

Across the four panels in Fig.~4, the main qualitative trend is that higher-order HTET corrections progressively reduce the residual \(E_{\max}\)-dependence of the extracted gaps. In particular,  compared to the raw truncation and the LO prescriptions, both NLO and NNLoc display a markedly flatter behavior as \(E_{\max}\) increases.

At low \(E_{\max}\), the NNLoc curves can display a stronger variation, in particular at larger coupling, where the NNLoc points approach the asymptotic region from below with a pronounced rise. However, as \(E_{\max}\) increases, the NNLoc results converge rapidly and become nearly indistinguishable from the NLO ones, reaching an approximately constant plateau already at intermediate cut-offs. In this sense, the NNLoc prescription does not improve the behavior at the smallest truncations, but it ensures faster convergence toward the large-\(E_{\max}\) regime and a comparably stable plateau once \(E_{\max}\) is sufficiently large.

\section{Conclusion}
\label{sec_Conclusion}
In this work we investigated HT for the two-dimensional \(\lambda\phi^{4}\) theory within the HTET framework, where truncation effects are organized as a controlled expansion in inverse powers of the UV cut-off \(E_{\max}\).

Building on the \(\mathcal{O}(V^{2})\) matching program of Ref.~\cite{Cohen2022}, we developed two complementary improvements.

First, we derived compact all-order expressions for the local matching corrections to the quartic coupling and the mass, obtained by resumming infinite classes of diagrams sharing a fixed topology within the local approximation.

Second, we extended the non-local sector beyond the leading terms by deriving the NNLoc operator insertions contributing at \(\mathcal{O}(E_{\max}^{-4})\), following and extending the systematic strategy of Ref.~\cite{Demiray:2025zqh}. These contributions are particularly relevant because, in view of a systematic \(\mathcal{O}(V^{3})\) HTET program, they scale with the same power of \(1/E_{\max}\) as the next local corrections.

On the numerical side, we quantified these improvements through the cut-off dependence of low-energy spectral gaps.
The resummed local prescription, implemented through Eqs.~\eqref{eq:delta_m_resum} and \eqref{eq:delta_lambda_resum}, does not yield a uniform improvement over the LO result in the explored range of \(E_{\max}\).
This behavior is consistent with the fact that the resummation captures only a restricted subset of UV contributions, a fixed topology treated in the local approximation, while non-local terms and additional topologies entering at the same order in the \(1/E_{\max}\) expansion can be numerically comparable to (or larger than) the higher-order local terms being resummed.
As a consequence, resumming this local subclass alone can lead to an overcorrection and to an apparently slower convergence with \(E_{\max}\) than the LO-improved Hamiltonian, which instead implements a controlled matching at fixed perturbative order and includes the complete \(\mathcal{O}(V^{2})\) local correction.

These observations motivated the inclusion of genuinely non-local operator insertions.
A central technical point is that at NNLoc the matching coefficients acquire distributional contributions involving derivatives of Heaviside step functions.
At finite volume, where the spectrum is discrete, introducing such distributions after discretization leads to an ambiguous numerical prescription.
We resolved this issue by determining the matching coefficients in infinite volume, where the distributional structure is unambiguously defined, and only then moved to finite volume to implement the operators on the discrete basis of free-Hamiltonian eigenstates, in the spirit of a continuum-first matching.
The resulting effective Hamiltonian combines continuum coefficients with finite-volume operators (summarized in Tab.~\ref{tab:operators_corrections}) and provides a systematic NNLoc extension of the non-local sector.

Our numerical results showed a clear trend: higher-order HTET corrections progressively reduce the residual \(E_{\max}\)-dependence of the extracted gaps.
Relative to the raw and LO truncation schemes, both NLO and NNLoc results display a markedly flatter behavior as \(E_{\max}\) increases.
At smaller cut-offs, NNLoc results can vary more strongly, especially at larger coupling, but they rapidly approach the large-\(E_{\max}\) regime and become nearly indistinguishable from NLO results at intermediate truncations, while maintaining a stable plateau once \(E_{\max}\) is sufficiently large.
For reproducibility, the spectra used to produce the figures are available in the accompanying database \cite{Github2026HTET}.

Overall, our analysis supports a consistent picture: a systematic HTET expansion beyond the leading local approximation is feasible, but its practical effectiveness depends on two closely related conditions.
From the analytical side, one must expand all previously derived non-local structures to the desired order in \(1/E_{\max}\), since subleading terms from lower-order non-local contributions can enter at the same parametric order as genuinely higher-order corrections.
From the numerical side, one must implement an increasingly richer operator basis, including higher-derivative structures associated with the irreducible dependence on external frequencies, and efficiently evaluate the corresponding matrix elements on the truncated Hilbert space.

In conclusion, a systematic HTET expansion must address both aspects, with the broader goal of improving precision at moderate truncation scales and of generalizing the framework to other quantum field theories and higher-dimensional settings.

\appendix
\section{Integral computation}
In this appendix we discuss the explicit evaluation of the integrals
\(F_n^\alpha\), \(I_n^\alpha\), and \(B_n^\alpha\), which enter the determination
of the matching coefficients. Their computation is technically non-trivial,
due to the presence of derivatives of the Heaviside step function and to the
non-linear dependence of its argument on the integration variables.

A first simplification follows from the observation that derivatives acting on
the integration variables can be equivalently transferred to derivatives with
respect to the energy cut-off. Considering \(F_n^\alpha\) as an example, one finds
\begin{equation}
    \label{eq:derivative_Theta}
    \frac{d^n}{d(2\omega(k))^n}\,
    \Theta\!\bigl(2\omega(k)-E_{\rm max}\bigr)
    =
    (-1)^n
    \frac{d^n}{dE_{\rm max}^n}\,
    \Theta\!\bigl(2\omega(k)-E_{\rm max}\bigr),
\end{equation}
which immediately implies
\begin{equation}
    \label{eq:Intagral_Fn_from_F0}
    F_n^\alpha(E_{\rm max})
    =
    (-1)^n
    \frac{d^n}{dE_{\rm max}^n}\,
    F_0^\alpha(E_{\rm max}) .
\end{equation}
Therefore, it is sufficient to compute \(F_0^\alpha(E_{\rm max})\) once and obtain all
higher-\(n\) functions by differentiation with respect to \(E_{\rm max}\).
The same simplification applies to the integrals \(I_n^\alpha\) and \(B_n^\alpha\),
where the derivatives with respect to \(W_3\) and \(W_4\) can be similarly
re-expressed as derivatives with respect to \(E_{\rm max}\).

A further simplification follows from the observation that the integrals scale
with the mass according to
\begin{equation}
\begin{aligned}
I^\alpha_n(E_{\max}; m)
&= m^{-\Delta_I}\,
I^\alpha_n\!\left(E;1\right),
&\Delta_I &= 1 + \alpha + n,
\\[4pt]
F^\alpha_n(E_{\max}; m)
&= m^{-\Delta_F}\,
F^\alpha_n\!\left(E;1\right),
&\Delta_F &= \alpha + n - 1,
\\[4pt]
B^\alpha_n(E_{\max}; m)
&= m^{-\Delta_B}\,
B^\alpha_n\!\left(E;1\right),
&\Delta_B &= 1 + \alpha + n ,
\end{aligned}
\end{equation}
with \(E \doteq E_{\rm max}/m\).
Therefore, without loss of generality, we can compute the integrals for \(m=1\)
and restore the mass dependence a posteriori.
In the case with four external legs, it is  sufficient to evaluate
\(F_0^\alpha(E,1)\), whose numerical computation is straightforward.

The situation is more involved in the case with two external legs, where one
has to determine \(I_0^\alpha(E,1)\).
Here one can proceed by noting that both \(P_3(k,q)\) and \(W_3(k,q)\), which from
now on we denote simply by \(P(k,q)\) and \(W(k,q)\), depend only on \(k^2\),
\(q^2\), and \((k+q)^2\).
This allows us to reduce the integral to
\[
I_0^\alpha(E) 
= 2\int_0^\infty \mathrm{d}k \int_{-\infty}^\infty \mathrm{d}q \; 
\frac{1}{P(k, q)W^\alpha(k, q)} \Theta(W(k, q) - E).
\]
Now we can decompose $I_0^\alpha(E)$ into three distinct contributions:
\begin{align}
\label{eq_I0alpha_regions}
I_0^\alpha(E) 
& =
I_{0, U}^\alpha(E) + I_{0, L}^{<,\alpha}(E) + I_{0, L}^{>,\alpha}(E), \\
\notag
I_{0, U}^\alpha(E)  &= 2\int_0^\infty \mathrm{d}k \int_{0}^\infty \mathrm{d}q \; 
\frac{1}{P(k, q)W^\alpha(k, q)} \Theta(W(k, q) - E),\\
\notag
I_{0, L}^{<,\alpha}(E) &= 2\int_0^\infty \mathrm{d}k \int_{0}^{k/2} \mathrm{d}q \; 
\frac{1}{P(k, -q)W^\alpha(k, -q)} \Theta(W(k, -q) - E), \\
\notag
I_{0, L}^{>,\alpha}(E) &= 2\int_0^\infty \mathrm{d}k \int_{k/2}^\infty \mathrm{d}q \; 
\frac{1}{P(k, -q)W^\alpha(k, -q)} \Theta(W(k, -q) - E),
\end{align}
where each individual integral is such that the function $W$ is monotonic with respect to $q$ at fixed $k$. We can therefore perform the change of integration variable $z = W(k, \pm q)$ at fixed $k$, obtaining:
\begin{align}
\label{eq_I0alpha_regions_W}
I_{0, U}^\alpha(E)  &= 2\int_0^\infty \mathrm{d}k \int_{W(k,0)}^\infty \frac{\mathrm{d}z}{z^\alpha} 
\frac{J(k,z)}{P(k, q_U(k, z))} \Theta(z - E),\\
I_{0, L}^{<,\alpha}(E) &= 2\int_0^\infty \mathrm{d}k \int_{W(k, -k/2)}^{W(k,0)} \frac{\mathrm{d}z}{z^\alpha}  
\frac{ J(k,z)}{P(k, q_L^<(k,z))} \Theta(z - E), \\
I_{0, L}^{>,\alpha}(E) &= 2\int_0^\infty \mathrm{d}k \int_{W(k,-k/2)}^\infty \frac{\mathrm{d}z}{z^\alpha}  
\frac{ J(k,z)}{P(k, q_L^>(k,z))} \Theta(z - E),
\end{align}
where we defined the inverse transformations as
\begin{alignat}{2}
q_U(k, z) &= -\frac{k}{2} + \sqrt{\frac{k^2}{4} + \frac{R^2(k,z)-k^2-1}{2(1+R(k,z))}} \qquad&&\text{ for } I_{0, U}^{\alpha}(E),\\
q_L^{<}(k, z) &= \frac{k}{2} - \sqrt{\frac{k^2}{4} + \frac{R^2(k,z)-k^2-1}{2(1+R(k,z))}}   \qquad&&\text{ for } I_{0, L}^{<,\alpha}(E),\\
q_L^{>}(k, z) &= \frac{k}{2} + \sqrt{\frac{k^2}{4} + \frac{R^2(k,z)-k^2-1}{2(1+R(k,z))}}  \qquad&&\text{ for } I_{0, L}^{>,\alpha}(E)  ,
\end{alignat}
with
\[
R(k, z) = \frac{z^2-1}{2} - z \omega(k), \quad
J(k, z) = \frac{d}{\mathrm{d}z} \qa 
\sqrt{\frac{k^2}{4} + \frac{R^2(k,z)-k^2-1}{2(1+R(k,z))}} \qc,
\]
being $J(k, z)$ the Jacobian of the variable transformation.

Now we can use the theta-function $\Theta(z-E)$ explicitly to change the integration limits.
In conclusion, we obtain
\begin{align}
\label{eq_I0alpha_regions_theta}
I_{0}^\alpha(E)  &= 2\int_0^\infty \mathrm{d}k \int_{\max(E,W(k,0))}^\infty \frac{\mathrm{d}z}{z^\alpha}
\frac{J(k,z)}{P(k, q_U(k, z))} \\
\notag
&+ 2\int_{k^*(E)}^\infty \mathrm{d}k \int_{\max(E,W(k, -k/2))}^{W(k,0)} \frac{\mathrm{d}z}{z^\alpha}
\frac{J(k,z)}{P(k, q_L^<(k,z))}  \\
\notag
&+ 2\int_0^\infty \mathrm{d}k \int_{\max(E,W(k,-k/2))}^\infty \frac{\mathrm{d}z}{z^\alpha} 
\frac{J(k,z)}{P(k, q_L^>(k,z))},
\end{align}
where we defined
\[
 k^*(E) \doteq \sqrt{\frac{(E-1)^2}{4}-1},
\]
which is the minimum value of $k$ such that $W(k, 0) > E $.

Now, the integral in the form above can be easily computed for any value of $E$ by means of a quadrature routine.
To take derivatives, we simply use polynomial interpolation, specifically we 
use the second barycentric form of the Chebyshev interpolant (see, for instance, Ref.~\cite{Diehl2021} and references therein)
\[
I_0^\alpha(E) = \sum_{j=0}^N I_0^\alpha(E(t_j)) b_j(t),
\]
where 
\begin{eqnarray*}
b_j(t) &=& \frac{\beta_j(-1)^j}{t-t_j} \Bigg/ \sum_{k=0}^{N}\frac{\beta_k(-1)^k}{t-t_k}, \quad\mbox{with}\,
\beta_j = 
\begin{cases}
1 & \text{ if } j\neq 0,N\\
1/2 & \text{ if } j=0,N
\end{cases},\\
E(t_j) &=& \frac{t_j+1}{2}(E_{\rm high} - E_{\rm low}) + E_{\rm low}.
\end{eqnarray*}
with  $E_{\rm high}, E_{\rm low}$ corresponding, respectively, to the maximum and minimum values
of $E$ for which we tabulate the integrals.

In this form the derivatives are trivially computed from
\[
\frac{d}{dE} I_0^\alpha(E) = \sum_{j=0}^N  b_j(t) \sum_{k=0}^N D_{jk} I_0^\alpha(E(t_k)),
\]
where the explicit expression for the derivative matrix $D$ 
can be found in section 2 of Ref.~\cite{Diehl2021}.
Repeated differentiation is achieved by applying the differentiation matrix multiple times.
A completely analogous strategy can be employed in the case with zero external legs, allowing for an efficient determination of the integral \(B_0^\alpha(E,1)\) as well.

\label{app:integral_computation}

\section{Summary of the leading non-local NLO corrections}

In this appendix we summarize the main result of Ref.~\cite{Demiray:2025zqh}, namely the leading non-local NLO corrections that enter the effective Hamiltonian at order \(\mathcal{O}(V^2)\) in the potential. 
Denoting this contribution by \(H_{\mathrm{eff}}^{\mathrm{(NLO)}}\), one finds
\begin{equation}
H_{\mathrm{eff}}^{\mathrm{(NLO)}} \;=\;
\xi\, H_0
+\frac{1}{2}\Bigl[
\alpha_1^{(1)}\,\{H_0,H_2\}+\alpha_2^{(2)}\,[H_0,H_2]
\Bigr]
+\frac{1}{24}\Bigl[
\beta_1^{(1)}\,\{H_0,H_4\}+\beta_2^{(2)}\,[H_0,H_4]
\Bigr] \, ,
\label{eq:NLO_nonlocal_summary}
\end{equation}
where the coefficients are given by
\begin{align}
\xi
&= -\frac{\lambda^{2}}{384(2\pi R)^{2}}
\sum_{1234}\delta_{1+2+3+4,0}\,
\frac{\delta(\omega_{1}+\omega_{2}+\omega_{3}+\omega_{4}-E_{\max})}
{\omega_{1}\omega_{2}\omega_{3}\omega_{4}(\omega_{1}+\omega_{2}+\omega_{3}+\omega_{4})}
\,, \notag \\[4pt]
\alpha_{1}^{(1)}
&= -\frac{\lambda^{2}}{48(2\pi R)^{2}}
\sum_{123}\delta_{1+2+3,0}\,
\frac{\delta(\omega_{1}+\omega_{2}+\omega_{3}-E_{\max})}
{\omega_{1}\omega_{2}\omega_{3}(\omega_{1}+\omega_{2}+\omega_{3})}
\,, \notag\\[4pt]
\alpha_{2}^{(2)}
&= -\frac{\lambda^{2}}{48(2\pi R)^{2}}
\sum_{123}\delta_{1+2+3,0}\,
\frac{\Theta(\omega_{1}+\omega_{2}+\omega_{3}-E_{\max})}
{\omega_{1}\omega_{2}\omega_{3}(\omega_{1}+\omega_{2}+\omega_{3})^{2}}
\,, \notag\\[4pt]
\beta_{1}^{(1)}
&= -\frac{3\lambda^{2}}{4\pi}\,
\frac{1}{E_{\max}^{2}\sqrt{E_{\max}^{2}-4m^{2}}}
\,, \notag\\[4pt]
\beta_{2}^{(2)}
&= -\frac{3\lambda^{2}}{64\pi R}\sum_{k}
\frac{\Theta(2\omega_{k}-E_{\max})}{\omega_{k}^{4}}
\,.
\end{align}
These expressions fully determine the NLO contribution to the effective Hamiltonian. 

\label{app:NLO}

\section{Finite-volume effects}
An interesting point to explore is the discrepancy between the correction coefficients
computed in the infinite volume case (i.e. continuum momentum space) 
and in the finite volume case (i.e. discrete momentum space).
As a case study, we can take the part of the NLO corrections to the coupling
that includes a delta function.
For reference, the correction is:
\[
\beta_1^{(1)} = -\frac{3\lambda^2}{32\pi R} 
   \sum_k \frac{\delta(2\omega_k-E_{\max})}{\omega_k^3}.
\]
The continuum version of the correction reads
\[
-\frac{3\lambda^2}{32\pi}F_1^3\doteq-\frac{3\lambda^2}{32\pi}\int_\mathbb{R} \mathrm{d}k\; 
   \frac{\delta(2\omega(k)-E_{\max})}{\omega(k)^3},
\]
in this case, the constraint imposed by the delta function, \(2\omega(k)=E_{\max}\), can be solved explicitly, allowing the integral to be evaluated in closed form and yielding
\begin{align}
\beta_1^{(1)} &= -\frac{3\lambda^2}{4\pi E_{\max}^2\sqrt{E_{\max}^2-4}}.
\end{align}
To obtain the correction in finite volume, without losing the appropriate 
treatment of the delta distribution, we can translate the correction
to position space. Therefore, starting from the continuum version of the correction, we can first take the Fourier transform to position space. In position space, after resolving all distributional dependencies, we can impose the finite volume. Finally, we transform back to momentum space to obtain a closed form in the now discrete momentum representation.
The main idea is to write
\[
\int_\mathbb{R} \mathrm{d}k \;f(k) \delta(2\omega(k)-E_{\max})
=\int_\mathbb{R} \mathrm{d}k \int \mathrm{d}x\; e^{-ikx} \tilde{f}(x)\delta(2\omega(k)-E_{\max}),
\]
where 
\[
\tilde{f}(x) = \int \frac{dq}{2\pi} e^{iqx} f(q) .
\]
In the specific case
\[
f(k) = \frac{1}{\omega(k)^3} \quad \Rightarrow\quad
\tilde{f}(x) = \frac{|x| K_1(|x|)}{\pi},
\]
where $K_1$ is the modified Bessel function of the second kind.

The delta function in $k$ imposes that $k=\pm k^*$, which we can use to obtain the following
representation
\begin{align}
\label{eq:beta1hat}
&-\frac{3\lambda^2}{32\pi} \frac{E_{\max}}{2\sqrt{E_{\max}^2-4}}
\int_{-\ell}^\ell \mathrm{d}x\; \frac{|x| K_1(|x|)}{\pi}\ta e^{-ik^*x}+e^{ik^*x}\tc
=-\frac{3\lambda^2}{4\pi E_{\max}^2\sqrt{E_{\max}^2-4}}
\underbrace{\qa
\frac{E_{\max}^3}{4\pi}\int_0^\ell \mathrm{d}x\; \cos(k^* x) \ x K_1(x)
\qc}_{\hat{\beta}_1^{(1)}},
\end{align}
where we have defined \(\ell \doteq \pi R\) and used the parity of the integrand to restrict the integration to positive values of \(x\).

Note that, in the infinite-volume limit \(\ell \to \infty\), the dimensionless factor \(\hat{\beta}_1^{(1)}\) approaches \(1\). This implies that finite-volume effects become negligible and the finite-volume expression for \(\beta_1^{(1)}\) reduces to the corresponding continuum result, consistently reproducing the correction computed in infinite volume.

\begin{table}[ht]
\centering
\begin{tabular}{c|c|c}
\hline
$E_{\max}$ & $\hat{\beta}_1^{(1)}(\ell = 5)$ & $\hat{\beta}_1^{(1)}(\ell = 10)$ \\
\hline
$10$ & $0.763252$  & $0.997042$ \\
$13$ & $1.28771$   & $1.00476$ \\
$15$ & $0.557404$  & $0.993781$ \\
$18$ & $1.6213$    & $1.00943$ \\
$20$ & $0.264269$  & $0.989269$ \\
$23$ & $2.03662$   & $1.01543$ \\
$25$ & $-0.145361$ & $0.983134$ \\
$28$ & $2.50951$   & $1.02268$ \\
$30$ & $-0.697936$ & $0.975158$\\
\hline
\end{tabular}
    \caption{Numerical values of the dimensionless factor \(\hat{\beta}_1^{(1)}\), defined in Eq.~\eqref{eq:beta1hat}, evaluated at fixed spatial extent \(\ell=\pi R\) for two representative volumes, \(\ell=5\) and \(\ell=10\), and for different values of the energy cut-off \(E_{\max}\).}
    \label{tab:finite_vol}
\end{table}
Table~\ref{tab:finite_vol} reports the numerical values of the finite-volume factor \(\hat{\beta}_1^{(1)}\) as a function of the energy cut-off \(E_{\max}\), for two representative choices of the spatial extent \(\ell\). The data illustrate how the finite-volume evaluation of the correction depends on both the volume and the UV cut-off.
For the default choice \(\ell = 5 \rightarrow R = 10/2\pi\), the difference between the finite-volume and infinite-volume results is sizable, and the results show a strong sensitivity to \(E_{\max}\).
In contrast, increasing to $\ell = 10 \Rightarrow R = 20 / 2\pi$ not only smooths 
the dependence on the energy cut-off but also  significantly reduces the 
discrepancy between the infinite volume and the finite volume results.
\FloatBarrier
\label{app:finite_volume}

\bibliography{Bibliography}

\end{document}